\documentclass[11pt]{article}
\usepackage{amsmath,amsfonts,amssymb,graphicx,epsfig,pdflscape}
\usepackage[usenames]{color}
\usepackage{pstricks}
\usepackage{arydshln,cite}
\numberwithin{equation}{section}

\definecolor{purple}{rgb}{1,0,1}
\definecolor{darkpurple}{rgb}{1,.2,1}
\definecolor{pink}{rgb}{1,.7,.7}

\newcommand{\nc}{\newcommand}
\nc\disp{\displaystyle}
\nc{\fh}{\hat{f}}
\nc{\muh}{\hat{\mu}}
\nc{\nuh}{\hat{\nu}}
\nc{\spos}[2]{\makebox(0,0)[#1]{$\sm{#2}$}}
\nc{\sm}[1]{{\scriptstyle #1}}
\nc{\qbar}{\overline{q}}
\nc{\bib}{\bibitem}
\nc{\al}{\alpha}
\nc{\g}{\gamma}
\nc{\G}{\Gamma}
\nc{\D}{\Delta}
\nc{\eps}{\epsilon}
\nc{\la}{\lambda}
\nc{\La}{\Lambda}
\nc{\var}{\varphi}
\nc{\pa}{\partial}
\nc{\nn}{\nonumber \\ }
\nc{\hf}{\frac{1}{2}}
\nc{\dz}{\frac{dz}{2\pi i}}
\nc{\bin}[2]{\left(\!\!\!\begin{array}{c} {#1}\\ {#2} \end{array}\!\!\!\right)}
\nc{\be}{\begin{equation}}
\nc{\ee}{\end{equation}}
\nc{\bea}{\begin{eqnarray}}
\nc{\eea}{\end{eqnarray}}
\nc{\bra}[1]{\langle {#1}|}
\nc{\ket}[1]{|{#1}\rangle}
\nc{\ketw}[1]{({#1})^{\phantom{a}}_{{\cal W}}}
\nc{\chit}{\raisebox{0.25ex}{$\chi$}}
\nc{\chih}{\raisebox{0.25ex}{$\hat\chi$}}
\nc{\Db}{\mbox{\boldmath $D$}}
\nc{\Hb}{\mbox{\boldmath $H$}}
\nc{\calH}{{\cal H}}
\nc{\calR}{{\cal R}}
\nc{\calL}{{\cal L}}
\nc{\calV}{{\cal V}}
\nc{\Hc}{{\cal H}}
\nc{\Rc}{{\cal R}}
\nc{\Lc}{{\cal L}}
\nc{\Vc}{{\cal V}}
\nc{\Ib}{\mbox{\boldmath $I$}}
\nc{\qb}{\bar{q}}
\nc{\Ac}{\mathcal{A}}
\nc{\Bc}{\mathcal{B}}
\nc{\Cc}{\mathcal{C}}
\nc{\Dc}{\mathcal{D}}
\nc{\Ec}{\mathcal{E}}
\nc{\Gc}{\mathcal{G}}
\nc{\Ic}{\mathcal{I}}
\nc{\Jc}{\mathcal{J}}
\nc{\Oc}{\mathcal{O}}
\nc{\Pc}{\mathcal{P}}
\nc{\Sc}{\mathcal{S}}
\nc{\Tc}{\mathcal{T}}
\nc{\Wc}{\mathcal{W}}
\nc{\Xc}{\mathcal{X}}
\nc{\Yc}{\mathcal{Y}}
\nc{\Zc}{\mathcal{Z}}
\nc{\fus}{\mbox{}\,\hat\otimes\,\mbox{}}
\nc{\Pch}{\hat{\Pc}}
\nc{\Rch}{\hat{\Rc}}
\nc{\Dh}{\hat{\Delta}}
\nc{\rh}{\hat{r}}
\nc{\sh}{\hat{s}}
\nc{\taub}{\bar{\tau}}
\nc{\Jcb}{\Jc_{\mathrm{b}}}
\nc{\rtt}{\mathtt{r}}
\nc{\stt}{\mathtt{s}}
\nc{\cosR}{\cos\frac{\pi p'rr'}{p}}
\nc{\cosS}{\cos\frac{\pi pss'}{p'}}
\nc{\sinR}{\sin\frac{\pi p'rr'}{p}}
\nc{\sinS}{\sin\frac{\pi pss'}{p'}}
\def\vvdots{\mathinner{\mkern1mu\raise1pt\vbox{\kern7pt\hbox{.}}\mkern2mu
  \raise4pt\hbox{.}\mkern2mu\raise7pt\hbox{.}\mkern1mu}}
\nc{\gauss}[2]{\left[\!\!\begin{array}{c} {#1}\\ {#2} \end{array}\!\!\right]}
\nc{\sbin}[2]{\left\{\!\!\!\begin{array}{c} {#1}\\ {#2} 
\end{array}\!\!\!\right\}}
\nc{\sbinlr}[2]{\Big\langle\!\!\begin{array}{c} {#1}\\ {#2} 
\end{array}\!\!\Big\rangle}
\nc{\bino}[2]{\left(\!\!\begin{array}{c} {#1}\\ {#2} \end{array}\!\!\right)}
\def\half {\mbox{$\textstyle \frac{1}{2}$}}
\def\vec#1{\mbox {\boldmath $#1$}}

\definecolor{lightblue}{rgb}{.61,.61,1}
\definecolor{midblue}{rgb}{.7,.7,1}
\definecolor{lightlightblue}{rgb}{.85,.85,1}
\definecolor{lightestblue}{rgb}{.96,.96,1}
\definecolor{lightpurple}{rgb}{1,.65,1}
\nc{\ch}{{\rm ch}}
\nc{\R}{{\cal R}}
\nc{\dkk}{\delta_{j,\{k,k'\}}^{(2)}}
\nc{\drr}{\delta_{j,\{r,r'\}}^{(2)}}
\nc{\ddkk}{\delta_{j,\{k,k'\}}^{(4)}}
\nc{\dddkk}{\delta_{j,\{k,k'\}}^{(8)}}
\nc{\dnn}{\delta_{j,\{n,n'\}}^{(2)}}
\nc{\ddnn}{\delta_{j,\{n,n'\}}^{(4)}}
\nc{\dddnn}{\delta_{j,\{n,n'\}}^{(8)}}

\definecolor{pink}{rgb}{1,.65,.65}

\begin{document}

\topmargin -5mm
\oddsidemargin 5mm

\setcounter{page}{1}

\vspace{8mm}
\begin{center}
{\huge {\bf Coset Graphs in Bulk and Boundary}}
\\[.3cm]
{\huge {\bf Logarithmic Minimal Models}}

\vspace{10mm}
{\Large Paul A. Pearce and J{\o}rgen Rasmussen}\\[.3cm]
{\em Department of Mathematics and Statistics, University of Melbourne}\\
{\em Parkville, Victoria 3010, Australia}\\[.4cm]
{\tt P.Pearce\,@\,ms.unimelb.edu.au}, \ \ \ {\tt J.Rasmussen\,@\,ms.unimelb.edu.au}

\end{center}

\vspace{8mm}
\centerline{{\bf{Abstract}}}
\vskip.4cm
\noindent
The logarithmic minimal models are not rational but, in the ${\cal W}$-extended picture, they resemble 
\mbox{rational} conformal field theories. 
We argue that the ${\cal W}$-projective representations are fundamental building blocks 
in both the boundary and bulk description of these theories. 
In the boundary theory, each ${\cal W}$-projective representation arising from fundamental fusion is associated with a boundary condition.
Multiplication in the associated Grothendieck ring leads to a Verlinde-like formula 
involving $A$-type twisted affine graphs $A^{(2)}_p$ and their coset graphs 
\mbox{$A^{(2)}_{p,p'}=A^{(2)}_p\otimes A^{(2)}_{p'}/{\mathbb Z}_2$}. This provides   
compact formulas for the conformal partition functions with $\Wc$-projective boundary conditions. 
On the torus, we propose modular invariant partition functions as sesquilinear forms in 
${\cal W}$-projective and 
rational minimal characters and observe that they are encoded by the same coset fusion graphs.  

%\\[.5cm]
%{\bf Keywords:}
%\\[.1cm]
%{\bf PACS number:}
%\end{titlepage}
%\newpage
\renewcommand{\thefootnote}{\arabic{footnote}}
\setcounter{footnote}{0}

\newpage
\tableofcontents

\newpage
\section{Introduction}

Logarithmic Conformal Field Theories (CFTs)~\cite{Gurarie} are relevant in describing many 
physical systems such as 
polymers~\cite{Saleur86,Saleur87a,Saleur87b,Duplantier86,DupSaleur86,Saleur92,PR06,Nigro09,PRV10}, 
percolation~\cite{Cardy92,Watts96,Cardy01,Smirnov01,MatRidout07,Ridout09,DubJacSal10}, 
symplectic fermions~\cite{Kausch95,Kausch00}, 
the Abelian sandpile model~\cite{Dhar90,Dhar99,MahRuelle01,Ruelle02,MRR0410,PGPR0710} 
and even minimal string theory~\cite{SeibergShih,IshimotoY}
and aspects of gravity~\cite{LSS0801,GJ0805,MSS0903,GJZ1010}. 
They also feature prominently in 
Wess-Zumino-Novikov-Witten models based on supergroups \cite{RS92,MS9605,Zir9905,SS0510,GQS0610,SS0611,CQS0708,Creutzig0908}, 
at fractional level \cite{Gab0105,LMRS0207,LMRS0311,Rid1001} or with affine 
Jordan blocks \cite{Ras0508},
as well as in vertex algebras \cite{AM0707,HLZ0710,Huang0712,NT0902,Huang09}.

The simplest logarithmic CFTs, the logarithmic minimal models, have their origin in the 
mid-nineties~\cite{Flohr96,GabKausch96,Flohr97}. Reviews of the status of the subject in 2001 
can be found in the papers~\cite{FlohrIran,GabIran} of the IPM School in Tehran~\cite{Tehran}. 
While further progress was made in the intervening years
\cite{FFHST0201,KN0203,FHST0306,Ras0405,Ras0408,GL0409,Nagi0504,FGST0504,Ras0507,FG0509,FGST0512}, 
the logarithmic minimal models came into sharp focus in 2006. 
In this year, these models were realized algebraically as logarithmic extensions of the minimal 
conformal field theories ${\cal M}(p,p')$ with a ${\cal W}_{p,p'}$ symmetry~\cite{FGST06} and, 
independently, as a series of exactly solvable lattice models ${\cal LM}(p,p')$~\cite{PRZ0607}. 
After much work, it is increasingly clear that these models in fact coincide. In particular, the 
representation content and chiral fusion algebras in both the Virasoro and the 
${\cal W}$-extended picture are quite well understood and agree in the 
lattice~\cite{RP0706,RP0707,PRR0803,RP0804,Rasm0805} and the algebraic 
approaches~\cite{EberleF06,RS0701,GabR08,Ras0812,GabRW09,Ras0906,Wood10,AM0908}. 
Also, in recent years, significant progress has been 
made~\cite{GR06,GabR08,GabRW09,GRW10} on the algebraic 
consistency of both the boundary and bulk CFTs for $c=-2$ and $c=0$, that is, ${\cal LM}(1,2)$ 
and ${\cal LM}(2,3)$ respectively.

In this paper, we denote the logarithmic minimal models by ${\cal WLM}(p,p')$ to emphasize 
that we are working in the ${\cal W}$-extended picture, not the Virasoro picture, and that we are assuming ${\cal W}_{p,p'}$ symmetry. These models have 
many properties in common with rational minimal models such as closure of fusion on certain
finite sets of chiral representations. Indeed, it is expected 
that these models will play a similar role for logarithmic CFT that the minimal models ${\cal M}(p,p')$~\cite{BPZ,ABF,FB} 
play for rational CFT. Rational theories, such as the minimal models, are classified~\cite{CIZ1,CIZ2,Kato} in the bulk by 
modular invariants~\cite{CardyModInv} associated with graphs leading to the famous $A$-$D$-$E$ classifications. 
Remarkably, the boundary rational theories are classified~\cite{BPPZ1,BPPZ2}, through nonnegative integer matrix 
representations (nimreps) of the Verlinde formula, by the same graphs. It is interesting to ask to what extent analogous statements hold in the context of logarithmic theories. 

In this paper, we explore connections between graphs and the bulk and boundary theories of the logarithmic 
minimal models.
In Section~\ref{SecConformal}, we consider the boundary theory and summarize the conformal 
data and relevant representation content of the logarithmic minimal models in the $\Wc$-extended 
picture. Among all possible representations, we emphasize $\Wc$-irreducible and $\Wc$-projective representations as the fundamental building 
blocks. Since all of these models share a common effective 
central charge $c^{\text{eff}}=1$, we recall some 
basic facts on affine $u(1)$ (Gaussian) theories in Section~\ref{SecBoundary}. 
Using the Dynkin diagrams of twisted affine Lie algebras 
$A_{p}^{(2)}$, we build the twisted affine coset graphs 
\mbox{$A^{(2)}_{p,p'}=A^{(2)}_p\otimes A^{(2)}_{p'}/{\mathbb Z}_2$} which we 
assert are classifying $A$-type graphs for ${\cal WLM}(p,p')$. Considering the modular 
transformations of the ${\cal W}$-projective representations, these graphs arise naturally from a 
Verlinde fusion algebra. Moving to the Grothendieck ring allows us to give a compact Verlinde-like 
formula for the boundary conformal partition functions for boundary conditions associated with the 
$\Wc$-projective representations. 
In this context, it is these $\Wc$-projective representations that fulfil this role.
For the logarithmic minimal models, the $\Wc$-irreducible 
representations are not in general associated 
with boundary conditions and can therefore not fulfil the role. 
In Section~\ref{SecBulk}, we consider the bulk theory. Extending the 
recent advance of \cite{GRW10} for ${\cal WLM}(2,3)$, we propose modular invariant partition 
functions for the 
general ${\cal WLM}(p,p')$ models and summarize our arguments in favour of this conclusion. 
In particular, we observe that the same classifying graphs encode the boundary and bulk theories. 
Additional arguments and some technical details are deferred to Appendix~\ref{AppEvidence}.
We finish with a brief discussion in Section~\ref{SecDiscussion} of some open problems.

\section{Conformal Data and Chiral Representation Content}
\label{SecConformal}

For rational CFTs, there are a finite number of irreducible representations which are the fundamental building blocks and 
close among themselves under fusion.
In the Virasoro picture, however, the logarithmic minimal models admit an infinite number of representations which close 
among themselves under fusion. 
To obtain a finite number of representations, we assume an extended ${\cal W}_{p,p'}$ symmetry~\cite{FGST06}. 
In this ${\cal W}$-extended picture, the infinity of Virasoro representations are reorganized into a finite number of 
${\cal W}$-indecomposable representations that close among themselves under fusion.
To emphasize the fact that we assume this extended symmetry, we use the notation ${\cal WLM}(p,p')$ even though these 
theories are described by the same lattice models, just with a more restricted class of boundary conditions that respect the 
${\cal W}$ symmetry.
For logarithmic CFTs, such as the logarithmic minimal models, we argue that the irreducible representations are 
supplemented with projective representations as the fundamental building blocks. The definition of a projective representation is given in Section~\ref{WprojSec}.

In this section, we review the relevant representation content of the logarithmic minimal models including 
%the Virasoro representations, 
the ${\cal W}$-irreducible representations, their projective covers and the 
projective Grothendieck 
generators. A summary  is given in Table~\ref{Wcontent} with the various Kac tables in 
Figures~\ref{VirKac}-\ref{Kacirredproj}. Details of our notation for these representations and their associated characters 
can be found, for example, in \cite{Rasm0805}. In the Kac tables (Figures~\ref{VirKac}-\ref{Kacirredproj}), we designate 
the $(r,s)$ entry as either corner, edge or interior (as indicated by shading) according to
\be
(r,s)=\begin{cases}
\mbox{corner},&r=0 \mbox{\;mod\;} p; \  s=0\mbox{\;mod\;}p'\\
\mbox{edge},&r\ne0\mbox{\;mod\;}p;\  s=0\mbox{\;mod\;}p'\mbox{\quad or\quad }r=0 \mbox{\;mod\;} p; \ 
 s\ne0\mbox{\;mod\;}p'\\
\mbox{interior},&r\ne0 \mbox{\;mod\;} p;\  s\ne0\mbox{\;mod\;}p'
\end{cases}
\ee
and associate to it a ``degree"
\be
d_{r,s}=\begin{cases}
1,&r=0 \mbox{\;mod\;} p; \  s=0\mbox{\;mod\;}p'\\
2,&r\ne0\mbox{\;mod\;}p;\  s=0\mbox{\;mod\;}p'\mbox{\quad or\quad }r=0 \mbox{\;mod\;} p; 
  \ s\ne0\mbox{\;mod\;}p'\\
4,&r\ne0 \mbox{\;mod\;} p;\  s\ne0\mbox{\;mod\;}p'
\end{cases}
\label{dimensions}
\ee
that is,
\be
 d_{r,s}=(2-\delta_{r,0}^{(p)})(2-\delta_{s,0}^{{(p')}}),
 \qquad \delta_{n,m}^{(N)}=\left\{\begin{array}{ll}1,\quad &n\equiv m\ (\mathrm{mod}\ N)\\[6pt]
    0,\quad &\mathrm{otherwise}  \end{array} \right.
\label{drs}
\ee

\subsection{Central charge and conformal weights}

The logarithmic minimal models ${\cal LM}(p,p')$ have central charges
\be
 c=1-\frac{6(p-p')^2}{pp'},\qquad 1\le p<p' \qquad (\mbox{$p, p'$ coprime integers})
\ee
In the Virasoro picture, there are an infinite number of so-called 
Kac representations \cite{PRZ0607,RP0707} with an infinitely extended Kac table of 
conformal weights
\be
 \Delta_{r,s}=\frac{(p'r-ps)^2-(p-p')^2}{4pp'},\qquad\qquad r,s\in\mathbb{N}
\ee
as indicated in Figure~\ref{VirKac}. We also use this Kac formula for conformal weights for {\em arbitrary} $r,s\in\mathbb{Z}$ 
and note the general $\mathbb{Z}_2$ Kac-table symmetry
\be
 \Delta_{r,s}=\Delta_{p-r,p'-s}
\label{Kacsym}
\ee
These models are nonunitary with a negative minimal conformal weight
\be
 \Delta_{\text{min}}=-\frac{(p-p')^2}{4pp'}
\label{Dmin}
\ee
The effective central charge and effective conformal weights are therefore
\be
 c^{\text{eff}}=c-24\Delta_{\text{min}}=1,\qquad 
  \Delta^{\text{eff}}_{r,s}=\Delta_{r,s}-\Delta_{\text{min}}=\frac{(p'r-ps)^2}{4pp'},
  \qquad r,s\in\mathbb{Z}
\ee

\begin{figure}[p]
{\vspace{0in}\psset{unit=1.cm}
{\small
\begin{center}
%Critical Percolation
\qquad\qquad
\begin{pspicture}(0,0)(7,11)
\psframe[linewidth=0pt,fillstyle=solid,fillcolor=lightestblue](0,0)(7,11)
%\psframe[linewidth=1pt,fillstyle=solid,fillcolor=lightestblue](0,0)(1,2)
\psframe[linewidth=0pt,fillstyle=solid,fillcolor=lightlightblue](1,0)(2,11)
\psframe[linewidth=0pt,fillstyle=solid,fillcolor=lightlightblue](3,0)(4,11)
\psframe[linewidth=0pt,fillstyle=solid,fillcolor=lightlightblue](5,0)(6,11)
\psframe[linewidth=0pt,fillstyle=solid,fillcolor=lightlightblue](0,2)(7,3)
\psframe[linewidth=0pt,fillstyle=solid,fillcolor=lightlightblue](0,5)(7,6)
\psframe[linewidth=0pt,fillstyle=solid,fillcolor=lightlightblue](0,8)(7,9)
\multiput(0,0)(0,3){3}{\multiput(0,0)(2,0){3}{\psframe[linewidth=0pt,fillstyle=solid,fillcolor=midblue](1,2)(2,3)}}
%\multirput(2,1)(2,0){3}{\pswedge[fillstyle=solid,fillcolor=red,linecolor=red](0,0){.25}{180}{270}}
%\multirput(2,2)(2,0){3}{\pswedge[fillstyle=solid,fillcolor=red,linecolor=red](0,0){.25}{180}{270}}
%\multirput(2,3)(2,0){3}{\pswedge[fillstyle=solid,fillcolor=red,linecolor=red](0,0){.25}{180}{270}}
%\multirput(1,3)(0,3){3}{\pswedge[fillstyle=solid,fillcolor=red,linecolor=red](0,0){.25}{180}{270}}
%\multirput(2,3)(0,3){3}{\pswedge[fillstyle=solid,fillcolor=red,linecolor=red](0,0){.25}{180}{270}}
\psgrid[gridlabels=0pt,subgriddiv=1]
\rput(.5,10.65){$\vdots$}\rput(1.5,10.65){$\vdots$}\rput(2.5,10.65){$\vdots$}
\rput(3.5,10.65){$\vdots$}\rput(4.5,10.65){$\vdots$}\rput(5.5,10.65){$\vdots$}
\rput(6.5,10.5){$\vvdots$}
\rput(.5,9.5){$12$}\rput(1.5,9.5){$\frac{65}8$}\rput(2.5,9.5){$5$}
\rput(3.5,9.5){$\frac{21}8$}\rput(4.5,9.5){$1$}\rput(5.5,9.5){$\frac{1}8$}\rput(6.5,9.5){$\cdots$}
\rput(.5,8.5){$\frac{28}3$}\rput(1.5,8.5){$\frac{143}{24}$}\rput(2.5,8.5){$\frac{10}3$}
\rput(3.5,8.5){$\frac{35}{24}$}\rput(4.5,8.5){$\frac 13$}\rput(5.5,8.5){$-\frac{1}{24}$}
\rput(6.5,8.5){$\cdots$}
\rput(.5,7.5){$7$}\rput(1.5,7.5){$\frac {33}8$}\rput(2.5,7.5){$2$}
\rput(3.5,7.5){$\frac{5}8$}\rput(4.5,7.5){$0$}\rput(5.5,7.5){$\frac{1}8$}
\rput(6.5,7.5){$\cdots$}
\rput(.5,6.5){$5$}\rput(1.5,6.5){$\frac {21}8$}\rput(2.5,6.5){$1$}
\rput(3.5,6.5){$\frac{1}8$}\rput(4.5,6.5){$0$}\rput(5.5,6.5){$\frac{5}8$}\rput(6.5,6.5){$\cdots$}
\rput(.5,5.5){$\frac{10}3$}\rput(1.5,5.5){$\frac {35}{24}$}\rput(2.5,5.5){$\frac 13$}
\rput(3.5,5.5){$-\frac{1}{24}$}\rput(4.5,5.5){$\frac 13$}\rput(5.5,5.5){$\frac{35}{24}$}
\rput(6.5,5.5){$\cdots$}
\rput(.5,4.5){$2$}\rput(1.5,4.5){$\frac 58$}\rput(2.5,4.5){$0$}\rput(3.5,4.5){$\frac{1}8$}
\rput(4.5,4.5){$1$}\rput(5.5,4.5){$\frac{21}8$}\rput(6.5,4.5){$\cdots$}
\rput(.5,3.5){$1$}\rput(1.5,3.5){$\frac 18$}\rput(2.5,3.5){$0$}\rput(3.5,3.5){$\frac{5}8$}
\rput(4.5,3.5){$2$}\rput(5.5,3.5){$\frac{33}8$}\rput(6.5,3.5){$\cdots$}
\rput(.5,2.5){$\frac 13$}\rput(1.5,2.5){$-\frac 1{24}$}\rput(2.5,2.5){$\frac 13$}
\rput(3.5,2.5){$\frac{35}{24}$}\rput(4.5,2.5){$\frac{10}3$}\rput(5.5,2.5){$\frac{143}{24}$}
\rput(6.5,2.5){$\cdots$}
\rput(.5,1.5){$0$}\rput(1.5,1.5){$\frac 18$}\rput(2.5,1.5){$1$}\rput(3.5,1.5){$\frac{21}8$}
\rput(4.5,1.5){$5$}\rput(5.5,1.5){$\frac{65}8$}\rput(6.5,1.5){$\cdots$}
\rput(.5,.5){$0$}\rput(1.5,.5){$\frac 58$}\rput(2.5,.5){$2$}\rput(3.5,.5){$\frac{33}8$}\rput(4.5,.5){$7$}
\rput(5.5,.5){$\frac{85}8$}\rput(6.5,.5){$\cdots$}
{\color{blue}
\rput(.5,-.5){$1$}
\rput(1.5,-.5){$2$}
\rput(2.5,-.5){$3$}
\rput(3.5,-.5){$4$}
\rput(4.5,-.5){$5$}
\rput(5.5,-.5){$6$}
\rput(6.5,-.5){$r$}
\rput(-.5,.5){$1$}
\rput(-.5,1.5){$2$}
\rput(-.5,2.5){$3$}
\rput(-.5,3.5){$4$}
\rput(-.5,4.5){$5$}
\rput(-.5,5.5){$6$}
\rput(-.5,6.5){$7$}
\rput(-.5,7.5){$8$}
\rput(-.5,8.5){$9$}
\rput(-.5,9.5){$10$}
\rput(-.5,10.5){$s$}}
\end{pspicture}
\end{center}}}
\caption{Infinitely extended Kac table of conformal weights for 
critical percolation ${\cal LM}(2,3)$. The corner, edge and interior entries are indicated by 
shading. 
\label{VirKac}}
\end{figure}
%%%%%%%%%%%%%%%%%%%%%%%%%%%%%%%%%%%%%%%%%%%%%%%
\begin{figure}[p]
\begin{center}
\psset{unit=1.cm}
\small
%Critical Percolation
\begin{pspicture}(-.5,-.7)(4,3)
\psframe[linewidth=0pt,fillstyle=solid,fillcolor=lightestblue](0,0)(4,3)
\psframe[linewidth=0pt,fillstyle=solid,fillcolor=lightlightblue](0,0)(1,3)
\psframe[linewidth=0pt,fillstyle=solid,fillcolor=lightlightblue](2,0)(3,3)
\psframe[linewidth=0pt,fillstyle=solid,fillcolor=lightlightblue](0,0)(4,1)
\multiput(0,0)(2,0){2}{\psframe[linewidth=0pt,fillstyle=solid,fillcolor=midblue](0,0)(1,1)}
\psgrid[gridlabels=0pt,subgriddiv=1](0,0)(4,3)
\rput(.5,2.5){$\frac 58$}\rput(1.5,2.5){$2$}\rput(2.5,2.5){$\frac {33}8$}\rput(3.5,2.5){$7$}
\rput(.5,1.5){$\frac 18$}\rput(1.5,1.5){$1$}\rput(2.5,1.5){$\frac{21}8$}\rput(3.5,1.5){$5$}
\rput(.5,.5){$-\frac{1}{24}$}\rput(1.5,.5){$\frac 13$}\rput(2.5,.5){$\frac{35}{24}$}
\rput(3.5,.5){$\frac {10}3$}
\rput(2,-1.1){\bf (a) ${\cal W}$-Irreducible}
{\color{blue}
\rput(.5,-.5){$0$}
\rput(1.5,-.5){$1$}
\rput(2.5,-.5){$2$}
\rput(3.5,-.5){$3$}
\rput(4.5,-.5){$\rh$}
\rput(-.5,.5){$0$}
\rput(-.5,1.5){$1$}
\rput(-.5,2.5){$2$}
\rput(-.5,3.5){$\sh$}}
\end{pspicture}
\hspace{1cm}
\begin{pspicture}(-.5,-.7)(4,4)
\psframe[linewidth=0pt,fillstyle=solid,fillcolor=lightestblue](0,0)(4,3)
\psframe[linewidth=0pt,fillstyle=solid,fillcolor=lightlightblue](0,0)(1,3)
\psframe[linewidth=0pt,fillstyle=solid,fillcolor=lightlightblue](2,0)(3,3)
\psframe[linewidth=0pt,fillstyle=solid,fillcolor=lightlightblue](0,0)(4,1)
\multiput(0,0)(2,0){2}{\psframe[linewidth=0pt,fillstyle=solid,fillcolor=midblue](0,0)(1,1)}
\psgrid[gridlabels=0pt,subgriddiv=1](0,0)(4,3)
\rput(.5,2.5){$\frac 58,\!\color{blue}\frac 58$}\rput(1.5,2.75){$0,\!2$}\rput(1.5,2.25){$0,\!\color{blue}2$}\rput(2.5,2.5){$\frac 18,\!\color{blue}\frac {33}8$}\rput(3.5,2.75){$0,\!1$}\rput(3.5,2.25){$2,\!\color{blue}7$}
\rput(.5,1.5){$\frac 18,\!\color{blue}\frac 18$}\rput(1.5,1.75){$0,\!1$}\rput(1.5,1.25){$0,\!\color{blue}1$}\rput(2.5,1.5){$\frac 58,\!\color{blue}\frac{21}8$}\rput(3.5,1.75){$0,\!1$}\rput(3.5,1.25){$2,\!\color{blue}5$}
\rput(.5,.5){$\color{blue}-\frac{1}{24}$}\rput(1.5,.5){$\frac 13,\!\color{blue}\frac 13$}\rput(2.5,.5){$\color{blue}\frac{35}{24}$}\rput(3.5,.5){$\frac 13,\!\color{blue}\frac {10}3$}
\rput(2,-1.1){\bf (b) Projective Covers}
{\color{blue}
\rput(.5,-.5){$0$}
\rput(1.5,-.5){$1$}
\rput(2.5,-.5){$2$}
\rput(3.5,-.5){$3$}
\rput(4.5,-.5){$\rh$}
\rput(-.5,.5){$0$}
\rput(-.5,1.5){$1$}
\rput(-.5,2.5){$2$}
\rput(-.5,3.5){$\sh$}}
\end{pspicture}
\hspace{.7cm}
%
%\hspace{2cm}
%
\begin{pspicture}(-1,-.7)(3,4.25)
\psframe[linewidth=0pt,fillstyle=solid,fillcolor=lightlightblue](0,0)(3,4)
 \psframe[linewidth=0pt,fillstyle=solid,fillcolor=lightestblue](1,1)(2,3)
 \multirput(0,0)(0,3){2}{\multirput(0,0)(2,0){2}{\psframe[linewidth=0pt,fillstyle=solid,fillcolor=midblue](0,0)(1,1)}}
 \psgrid[gridlabels=0pt,subgriddiv=1](0,0)(3,4)
 \rput(.5,3.5){$\frac{35}{24}$}
 \rput(1.5,3.5){$\frac{1}{3}$}
 \rput(2.5,3.5){$-\frac{1}{24}$}
  \rput(.5,2.5){$\frac{5}{8}$}
 \rput(1.5,2.5){$0$}
 \rput(2.5,2.5){$\frac{1}{8}$}
 \rput(.5,1.5){$\frac{1}{8}$}
 \rput(1.5,1.5){$0$}
  \rput(2.5,1.5){$\frac{5}{8}$}
 \rput(.5,.5){$-\frac{1}{24}$}
 \rput(1.5,.5){$\frac{1}{3}$}
  \rput(2.5,.5){$\frac{35}{24}$}
  \rput(1.5,-1.1){\bf (c) Proj Grothendieck}
{\color{blue}
 \rput(.5,-.5){$0$}
 \rput(1.5,-.5){$1$}
 \rput(2.5,-.5){$2$}
 \rput(3.5,-.5){$r$}
 \rput(-.5,.5){$0$}
 \rput(-.5,1.5){$1$}
 \rput(-.5,2.5){$2$}
  \rput(-.5,3.5){$3$}
  \rput(-.5,4.5){$s$}}
\end{pspicture}\quad\mbox{}
\end{center}
\caption{Kac tables of conformal weights for critical percolation ${\cal WLM}(2,3)$: 
(a) the $2pp'=12$ \mbox{$\Wc$-irreducible} representations ${\cal W}(\hat\Delta_{\rh,\sh})$,
(b) the $2pp'=12$ \mbox{$\Wc$-projective} representations $\Pch_{\rh,\sh}$
and 
(c) the $\half (p+1)(p'+1)=6$ projective Grothendieck generators ${\cal G}_{r,s}$. 
The corner, edge and interior entries are indicated by shading. 
The Kac table in (b) organizes the ${\cal W}$-projective representations $\Pch_{\rh,\sh}$
in a one-to-one correspondence with the Kac table for the (non-minimal) $\Wc$-irreducible representations
in table (a) and (\ref{irredrs}). The conformal weights given in table (b) are the ones appearing in (\ref{PRD}).
In each position $(\rh,\sh)$ in table (b), the greatest conformal weight is identical to the corresponding
conformal weight in table (a).
The usual rational Kac table of the minimal ${\cal W}$-irreducible representations is not shown
by itself as it is identical to the interior part of the projective Grothendieck table (c).
\label{Kacirredproj}}
\end{figure}

\begin{table}[t]
\begin{center}
$
\renewcommand{\arraystretch}{1.2}
\begin{array}{|c||c|c|c|c|c|c|}
\hline
&\mbox{Reps}&\mbox{Chars}&\mbox{b.c.}&\mbox{Number\rule{0pt}{16pt}}&{\mbox{Symp} \atop \mbox{\rule{0pt}{8pt}Ferm}}&{\mbox{Crit}\atop\mbox{\rule{0pt}{8pt}Perc}}\\[4pt]
\hline \hline
\mbox{${\cal W}$-irred reps}&\Wc(\D)&\chit[\Wc(\D)](q)&\times&2pp'+\half(p\!-\!1)(p'\!-\!1)&4&13
 \\[0pt]
\hline
\mbox{Corner}&\Wc(\D_{\kappa p,p'})&\chit_{\kappa p,p'}(q)&\checkmark&2&2&2\\[0pt]
\mbox{Edge}&\Wc(\D_{a,\kappa p'})&\chit_{a, \kappa p'}(q)&\checkmark&2(p-1)&0&2\\[0pt]
%\hline
\mbox{Edge}&\Wc(\D_{\kappa p,b})&\chit_{\kappa p,b}(q)&\checkmark&2(p'-1)&2&4\\[0pt]
%\hline
\mbox{Interior}&\Wc(\D_{\kappa p\!+\!a,b})&\chit_{\kappa p+a,b}(q)&\times&2(p-1)(p'-1)&0&4
\\[1pt]
\hdashline
\mbox{Minimal}&\Wc(\D_{a,b})&\mbox{ch}_{a,b}(q)&\times&\half(p-1)(p'-1)&0&1\\[0pt]
\hline\hline
\mbox{Proj covers}&\Pc(\D)={\hat{\cal R}_{\kappa p,p'}^{r,s}}&\chi[{\hat{\cal R}_{\kappa p,p'}^{r,s}}](q)&\times&
 2pp'+\half(p\!-\!1)(p'\!-\!1)&4&13
\\[2pt]
\hline
\mbox{Rank 1}&{\hat{\cal R}_{\kappa p,p'}^{0,0}}\equiv\Wc(\D_{\kappa p,p'})&\chi[{\hat{\cal R}_{\kappa p,p'}^{0,0}}](q)&\checkmark&2&2&2
\\[3pt]
\mbox{Rank 2}&\Rch_{\kappa p,p'}^{p-a,0}&\chit[\Rch_{\kappa p,p'}^{p-a,0}](q)&\checkmark&2(p-1)&0&2
\\[3pt]
\mbox{Rank 2}&\Rch_{p,\kappa p'}^{0,p'-b}&\chit[\Rch_{p,\kappa p'}^{0,p'-b}](q)&\checkmark&2(p'-1)&2&4
\\[3pt]
\mbox{Rank 3}&\Rch_{\kappa p,p'}^{a,p'-b}&\chit[\Rch_{\kappa p,p'}^{a,p'-b}](q)&\checkmark&2(p-1)(p'-1)&0&4
\\[2pt]
\hdashline
\mbox{Min proj covers}&{\cal P}_{a,b}&\chit[{\cal P}_{a,b}](q)&\times&\half(p-1)(p'-1)&0&1\\
\hline\hline
\mbox{Proj Groth}&\Gc_{r,s}&d_{r,s}\varkappa^n_{r,s}(q)&(\checkmark)&\half(p+1)(p'+1)&3&6\\
%\mbox{${\cal W}$-fund reps}&\hat{\cal R}^{r,s}_{\kappa p,\kappa' p'}&\chi[\hat{\cal R}^{r,s}_{\kappa p,\kappa' p'}](q)&\checkmark&6pp'-2p-2p'&6&26\\[0pt]
%\hline
%\mbox{${\cal W}$-proj reps}&{\hat{\cal R}_{\kappa p,p'}^{r,s}}&\chi[\hat{\cal R}^{r,s}_{\kappa p,p'}](q)&\checkmark&2pp'&4&12\\[2pt]
\hline
%\mbox{Min reps}&\mbox{Vir}_{r,s}&\mbox{ch}_{r,s}(q)&\times&(p-1)(p'-1)&0&2\\
%\hline
\hline
\end{array}
$
\end{center}
\caption{Representation content of ${\cal WLM}(p,p')$ with the notation 
$\kappa=1,2$; $a=1,2,\ldots,p-1$ and \mbox{$b=1,2,\ldots, p'-1$}. 
Each ${\cal W}$-irreducible representation admits a unique projective cover.
To every ${\cal W}$-irreducible representation listed explicitly,
the corresponding projective cover is also listed:
$\Pc(\D_{a,\kappa p'})=\Rch_{\kappa p,p'}^{p-a,0}$, for example. 
As indicated in the column denoted by b.c., 
the \mbox{${\cal W}$-irreducible} representations and their projective covers 
do not in general have associated boundary conditions. 
The projective Grothendieck generators are listed here for convenience
with the notation $r=0,1,\ldots,p$ and $s=0,1,\ldots,p'$. They are not
representations and do not have conjugate boundary conditions, but each representative
of the equivalence class defining a given generator does.
The counting of representations and projective Grothendieck generators is given 
explicitly for symplectic fermions ${\cal WLM}(1,2)$ and critical percolation ${\cal WLM}(2,3)$.
\label{Wcontent}}
\end{table}

\subsection{${\cal W}$-irreducible representations}

The ${\cal W}$-irreducible representations are the fundamental building blocks respecting 
${\cal W}_{p,p'}$ symmetry. The ${\cal W}_{p,p'}$ algebra is generated by
the energy-momentum tensor $T(z)$ and two Virasoro primaries $W^+(z)$ and $W^-(z)$ of
conformal dimension $(2p-1)(2p'-1)$.

The $2pp'+\tfrac{1}{2}(p-1)(p'-1)$ 
${\cal W}$-irreducible representations $\Wc(\D)$ of ${\cal WLM}(p,p')$ are 
uniquely determined by their conformal weights $\D$. 
There are $\half (p-1)(p'-1)$ ${\cal W}$-irreducible representations corresponding to the 
representations of the rational minimal models with conformal weights
\be
 \Delta^{\text{Min}}_{r,s}=\Delta_{r,s},
  \quad\qquad 1\le r \le p-1\ ,\quad 1\le s \le p'-1
\label{irredMin}
\ee
These are organized into the usual Kac table with the ${\mathbb Z}_2$ Kac-table symmetry (\ref{Kacsym}) 
with associated characters given by
\be
 \ch_{r,s}(q)=\chit[\Wc(\D_{r,s})](q)
  =\frac{1}{\eta(q)} \sum_{k\in{\mathbb Z}} 
   \big(q^{(rp'-sp+2kpp')^2/4pp'}-q^{(rp'+sp+2kpp')^2/4pp'}\big)
\label{ch}
\ee
where the Dedekind eta function is
\be
 \eta(q)=q^{1/24} \prod_{k=1}^\infty(1-q^k),\qquad q=e^{2\pi i\tau}
\ee
The remaining $2pp'$ (non-minimal) ${\cal W}$-irreducible representations can be organized into a Kac table, as in 
Figure~\ref{Kacirredproj}, with conformal weights
\be
 \hat\Delta_{\rh,\sh}=\Delta_{p+\rh,p'-\sh},\qquad 0\le \rh\le 2p-1,\quad 0\le \sh\le p'-1
\label{irredrs}
\ee
and the caveat that there is no Kac-table symmetry here since the entries are all distinct. 
The associated $\Wc$-irreducible characters can be written in a uniform way as
\bea
 \chih_{\rh,\sh}(q)&=&\chit_{p+\rh,p'-\sh}(q)=\chit[\Wc(\hat{\D}_{\rh,\sh})](q)\nn
  &=&\frac{1}{\eta(q)} \sum_{k\in{\mathbb Z}} k(k+\lfloor\tfrac{\rh}{p}\rfloor)
   \big(q^{(\rh p'+\sh p+2(k-1)pp')^2/4pp'}-q^{(\rh p'-\sh p+2kpp')^2/4pp'}\big)
\label{chih}
\eea

\subsection{${\cal W}$-projective representations and their fusion algebra}
\label{WprojSec}

Loosely speaking, a $\Wc$-projective representation is a `maximal $\Wc$-indecomposable'
representation in the sense that it does not appear as a subfactor of any $\Wc$-indecomposable
representation different from itself. Likewise in a loose sense, a projective cover of a
$\Wc$-indecomposable representation $\Rch$ is a $\Wc$-projective representation 
containing $\Rch$ as a quotient.
Precise definitions are given as follows.

The sequence of modules
\be
 0\to A\to B\to C\to 0
\label{short}
\ee
is a short exact sequence if $A$ is mapped injectively into $B$ which is
mapped surjectively onto $C$, with the kernel of the latter map being the image
of the former map.
The short exact sequence (\ref{short}) is split if $B\simeq A\oplus C$.
The module $P$ is projective if all short exact sequences
\be
 0\to A\to B\to P\to 0
\label{proj}
\ee
are split. 
A module $P$ is a projective cover of the module $M$ if $P$ is a projective
module and there exists an epimorphism (onto or surjective homomorphism) from $P$ to $M$ whose kernel
is a superfluous submodule of $P$. Here a submodule $S$ of a module $B$ is
superfluous or inessential if for every submodule $C$ of $B$, $C+S=B$ implies $C=B$. 
In particular, the zero module is a superfluous submodule of any module.
A projective cover of a module does not always exist, but when it does it is unique
up to isomorphisms.

There are $2pp'$ ${\cal W}$-projective representations associated with
boundary conditions in ${\cal WLM}(p,p')$ \cite{Rasm0805} and we denote them by
\be
 \Rch_{\kappa p,\kappa'p'}^{r,s},\qquad \kappa,\kappa'=1,2;\quad 
   0\leq r\leq p-1;\quad 0\leq s\leq p'-1
\label{R}
\ee
This notation assumes that 
\be
 \Rch_{2p,p'}^{r,s}=\Rch_{p,2p'}^{r,s},\quad
 \Rch_{2p,2p'}^{r,s}=\Rch_{p,p'}^{r,s},\quad
 \Rch_{p,p'}^{0,0}=\Wc(\D_{p,p'}),\quad
 \Rch_{2p,p'}^{0,0}=\Wc(\D_{2p,p'})
\ee
and for later convenience, we extend it by 
\be
 \Rch_{p,p'}^{p,s}\equiv\Rch_{p,2p'}^{0,s},\qquad
 \Rch_{p,p'}^{r,p'}\equiv\Rch_{2p,p'}^{r,0},\qquad
 \Rch_{p,p'}^{p,p'}\equiv\Rch_{2p,2p'}^{0,0}\equiv\Rch_{p,p'}^{0,0}
\label{Rp}
\ee
The rank of a ${\cal W}$-projective representation (\ref{R}) is given by
\be
\mbox{rank}(\Rch_{\kappa p,\kappa'p'}^{r,s})=d_{r,s}-\lfloor\tfrac{d_{r,s}}{4}\rfloor
\ee
where the degree $d_{r,s}$ is defined by (\ref{dimensions}). We recall that the rank of a ${\cal W}$-indecomposable 
representation is the size of the largest Jordan block appearing in the Virasoro dilatation generator $L_0$.

Since a projective cover of a module $M$ must contain $M$ as a quotient, the 
projective covers of the ${\cal W}$-irreducible representations (\ref{irredrs})
follow immediately from the conjectured embedding patterns
for the rank-2 and rank-3 representations in \cite{Rasm0805}. 
In accordance with \cite{Wood10}, we thus have
\be
 \Pc(\D_{2p-r,s})=\Rch_{p,p'}^{p-r,p'-s},\qquad
 \Pc(\D_{3p-r,s})=\Rch_{2p,p'}^{p-r,p'-s},\qquad 1\leq r\leq p;\ 1\leq s\leq p'
\ee
and it follows that the two $\Wc$-irreducible
representations $\Wc(\D_{p,p'})$ and $\Wc(\D_{2p,p'})$ are their 
own projective covers. It is convenient to introduce the notation
\be
 \Pch_{\rh,\sh}=\Pc(\hat{\Delta}_{\rh,\sh})
   =\begin{cases}\hat{\cal R}^{\rh,\sh}_{p,p'},&\ 0\le \rh\le p-1,\ \phantom{2}0\le \sh\le p'-1\\[7pt]
\hat{\cal R}^{\rh-p,\sh}_{2p,p'},&\ p\le \rh\le 2p-1,\ 0\le \sh\le p'-1\end{cases}
\label{Wprojreps}
\ee
This association organizes the $2pp'$ ${\cal W}$-projective representations into a Kac table in 
one-to-one correspondence with the (non-minimal) ${\cal W}$-irreducible representations,
see Figure~\ref{Kacirredproj}. 
In this Kac table, the ${\cal W}$-projective representations are of rank 1, 2 and 3 on the corner, edge 
and interior entries respectively, that is, 
\be
 d_{\rh,\sh}=2^{\mathrm{rank}(\Pch_{\rh,\sh})-1},\qquad 0\le \rh\le 2p-1,\quad 0\le \sh\le p'-1
\label{ranks}
\ee
We note also that the dimensions of the ${\cal W}$-irreducible $\mbox{Hom}$ spaces have been 
conjectured~\cite{Wood10} to be 
\be
 \mathrm{dim}(\mathrm{Hom}(\Pch_{\rh,\sh},\Pch_{\rh,\sh}))=d_{\rh,\sh}
 \label{HomSpaces}
\ee
These dimensions will play a special role and are related to the Coxeter exponents of the coset graphs.

Depending on its rank, a $\Wc$-indecomposable representation is uniquely characterized by
1, 2 or 4 conformal weights \cite{RP0707,Rasm0805}.
A $\Wc$-projective representation $\Pch_{\rh,\sh}$, in particular, can thus be characterized
by $d_{\rh,\sh}$ conformal weights as follows
\be
 \begin{array}{rll}
 &\Pch_{0,0}=\Rch(\Dh_{0,0}),\qquad 
 &\Pch_{p,0}=\Rch(\Dh_{p,0})
\\[8pt]
 &\Pch_{a,0}=\Rch(\Dh_{a,0},\Dh_{a,0}),\qquad 
 &\Pch_{p+a,0}=\Rch(\Dh_{p-a,0},\Dh_{p+a,0}) 
\\[8pt]
 &\Pch_{0,b}=\Rch(\Dh_{0,b},\Dh_{0,b}),\qquad 
 &\Pch_{p,b}=\Rch(\Dh_{0,p'-b},\Dh_{p,b})\qquad 
\\[8pt]
 &\Pch_{a,b}=\Rch(\D_{a,b},\D_{a,b};\Dh_{a,b},\Dh_{a,b}),\qquad 
 &\Pch_{p+a,b}=\Rch(\D_{p-a,b};\Dh_{p-a,b},\Dh_{a,p'-b},\Dh_{p+a,b})
 \end{array}
\label{PRD}
\ee
where $1\leq a\leq p-1$ and $1\leq b\leq p'-1$. 
Within a given projective cover, the conformal weights differ by integers. In each case, 
the greatest conformal weight appears last and has the same labels as $\Pch_{\rh,\sh}$.
A semicolon is used to separate the conformal weights of the form (\ref{irredMin})
from the ones of the form (\ref{irredrs}). Due to the repetition of some conformal weights, this 
notation can obviously be simplified.
For ${\cal WLM}(2,3)$, it coincides with the notation used in \cite{GabRW09}.

According to the conjectured fusion rules of \cite{Rasm0805}, the ${\cal W}$-projective representations form a 
closed fusion algebra whose fusion rules can be written compactly as
\bea
 \!\!\!\!\!\!\Rch_{\kappa p,p'}^{r,s}\otimes\Rch_{p,\kappa'p'}^{r',s'}
 &=&\frac{d_{r,s}d_{r',s'}}{4}
 \!\left(
  \Big\{\bigoplus_{r''}^{p-|r-r'|-1}
      \oplus\bigoplus_{r''}^{|p-r-r'|-1}\Big\}
  \Big\{\bigoplus_{s''}^{p'-|s-s'|-1}
     \oplus\bigoplus_{s''}^{|p'-s-s'|-1}\Big\}
   \Rch_{\kappa p,\kappa'p'}^{r'',s''}
 \right.
 \nn
 &&\qquad\oplus\
  \Big\{\bigoplus_{r''}^{p-|p-r-r'|-1}
     \oplus\bigoplus_{r''}^{|r-r'|-1}\Big\}
  \Big\{\bigoplus_{s''}^{p'-|p'-s-s'|-1}
     \oplus\bigoplus_{s''}^{|s-s'|-1}\Big\}
  \Rch_{\kappa p,\kappa'p'}^{r'',s''}
 \nn
 &&\qquad\oplus\
  \Big\{\bigoplus_{r''}^{p-|r-r'|-1}
    \oplus\bigoplus_{r''}^{|p-r-r'|-1}\Big\}
  \Big\{\bigoplus_{s''}^{p'-|p'-s-s'|-1}
    \oplus\bigoplus_{s''}^{|s-s'|-1}\Big\}
  \Rch_{\kappa p,(2\cdot\kappa')p'}^{r'',s''}
 \nn
 &&\qquad\oplus\,   \left.
  \Big\{\bigoplus_{r''}^{p-|p-r-r'|-1}
     \oplus\bigoplus_{r''}^{|r-r'|-1}\Big\}
  \Big\{\bigoplus_{s''}^{p'-|s-s'|-1}
     \oplus\bigoplus_{s''}^{|p'-s-s'|-1}\Big\}
  \Rch_{\kappa p,(2\cdot\kappa')p'}^{r'',s''}
 \right)
\label{FusProj}
\eea
where we have introduced the shorthand notations $2\cdot1=2$, $2\cdot2=1$ and
\be
 \bigoplus_n^N R_n=\bigoplus_{n=\eps(N),\,\mathrm{by}\,2}^N R_n,\qquad 
 \eps(N)=\tfrac{1}{2}(1-(-1)^N)=N\ (\mathrm{mod}\ 2)
 \label{parity}
\ee
This fusion algebra does not contain an identity but is both associative and commutative and, despite appearances, 
the multiplicities are all nonnegative integers.

\subsection{Projective covers of minimal ${\cal W}$-irreducible representation}
\label{SecProjCover}

Just as the non-minimal ${\cal W}$-irreducible representations (\ref{irredrs}) have projective covers (\ref{Wprojreps}),
it seems natural to expect \cite{HLZ0710,Huang0712,NT0902,Huang09}
that every minimal ${\cal W}$-irreducible representation $\Wc(\D_{a,b})$ (\ref{irredMin}) also admits a projective cover
$\Pc_{a,b}=\Pc(\D_{a,b})$.
Strong evidence in favour of this for ${\cal WLM}(2,3)$ was recently provided in \cite{GRW10}.
Adopting their graphical description, we extend their arguments to ${\cal WLM}(p,p')$ by conjecturing that 
the embedding diagram of $\Pc_{a,b}$ for $1\leq a\leq p-1$ and $1\leq b\leq p'-1$ is given by
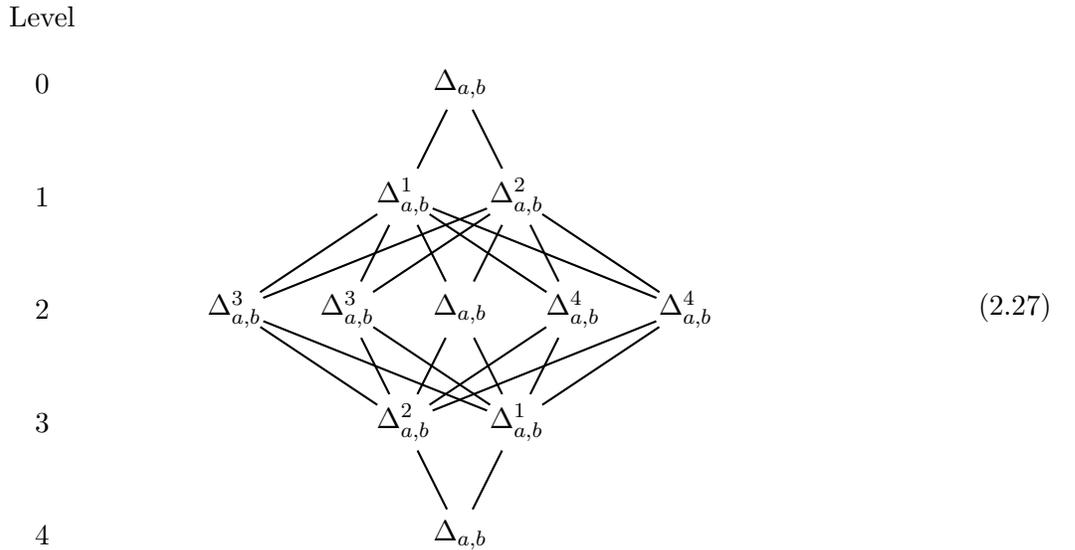
\bea
\psset{unit=1.5cm}
\begin{pspicture}[shift=-2.45](-0.4,0.5)(5,5.8)
%\psgrid
\psline(3,5)(2.5,4)
\psline(3,5)(3.5,4)
\psline(2.5,4)(1,3)
\psline(2.5,4)(2,3)
\psline(2.5,4)(3,3)
\psline(2.5,4)(4,3)
\psline(2.5,4)(5,3)
\psline(3.5,4)(1,3)
\psline(3.5,4)(2,3)
\psline(3.5,4)(3,3)
\psline(3.5,4)(4,3)
\psline(3.5,4)(5,3)
\psline(1,3)(2.5,2)
\psline(2,3)(2.5,2)
\psline(3,3)(2.5,2)
\psline(4,3)(2.5,2)
\psline(5,3)(2.5,2)
\psline(1,3)(3.5,2)
\psline(2,3)(3.5,2)
\psline(3,3)(3.5,2)
\psline(4,3)(3.5,2)
\psline(5,3)(3.5,2)
\psline(2.5,2)(3,1)
\psline(3.5,2)(3,1)
\pscircle[linewidth=0pt,linecolor=white,fillstyle=solid,fillcolor=white](3,5){.25}
\pscircle[linewidth=0pt,linecolor=white,fillstyle=solid,fillcolor=white](3,1){.25}
\multirput(-.5,1)(1,0){2}{\pscircle[linewidth=0pt,linecolor=white,fillstyle=solid,fillcolor=white](3,1){.275}}
\multirput(-2,2)(1,0){5}{\pscircle[linewidth=0pt,linecolor=white,fillstyle=solid,fillcolor=white](3,1){.275}}
\multirput(-.5,3)(1,0){2}{\pscircle[linewidth=0pt,linecolor=white,fillstyle=solid,fillcolor=white](3,1){.275}}
\rput(-0.7,5.6){Level}
%\rput(3,6){$\Pc_{a,b}$}
\rput(-0.7,5){$0$}
\rput(3,5){$\D_{a,b}$}
\rput(-0.7,4){$1$}
\rput(2.5,4){$\D_{a,b}^1$}
\rput(3.5,4){$\D_{a,b}^2$}
\rput(-0.7,3){$2$}
\rput(1,3){$\D_{a,b}^3$}
\rput(2,3){$\D_{a,b}^3$}
\rput(3,3){$\D_{a,b}$}
\rput(4,3){$\D_{a,b}^4$}
\rput(5,3){$\D_{a,b}^4$}
\rput(-0.7,2){$3$}
\rput(2.5,2){$\D_{a,b}^2$}
\rput(3.5,2){$\D_{a,b}^1$}
\rput(-0.7,1){$4$}
\rput(3,1){$\D_{a,b}$}
\end{pspicture}
\eea
where
\be
 \D_{a,b}^1=\D_{p+a,p'-b},\qquad
 \D_{a,b}^2=\D_{2p-a,b},\qquad
 \D_{a,b}^3=\D_{3p-a,p'-b},\qquad
 \D_{a,b}^4=\D_{2p+a,b}
\ee
The various $\Wc$-irreducible subfactors $\Wc(\D)$ are indicated by their respective 
conformal weights $\D$.
As in the description of the specific embedding diagrams in \cite{GRW10}, the action of a mode in $\Wc_{p,p'}$ on a 
vector in $\Wc(\D)$ gives a vector in the direct sum of the given $\Wc(\D)$ and all the subfactors joined to it by 
downward-directed lines (in one or more steps).
The corresponding character is given by
\bea
 \chit[\Pc_{a,b}](q)&=&3\,\ch_{a,b}(q)+2\big[\chit_{p+a,p'-b}(q)+\chit_{2p-a,b}(q)+\chit_{3p-a,p'-b}(q)
   +\chit_{2p+a,b}(q)\big]\nn
 &=&2\,\ch_{a,b}(q)+\tfrac{1}{2}\chit[\Rch_{p,p'}^{a,b}](q)
\label{Pchar}
\eea
where the character $\chit[\Rch_{p,p'}^{a,b}](q)$ of the $\Wc$-projective 
rank-3 representation $\Rch_{p,p'}^{a,b}$ is discussed in the following.

\subsection{Projective Grothendieck ring and Kac Table}
\label{SecProjKac}

We refer to the characters of the ${\cal W}$-projective representations as $\Wc$-projective characters.
Following \cite{Rasm0805}, these characters can be written as
\be
 \begin{array}{rcl}
 \chi\big[{\hat\R_{(\ell+1)p,p'}^{0,0}}\big](q)
   &=&\disp\frac{1}{\eta(q)}\sum_{k\in\mathbb{Z}}q^{(2k+\ell)^2pp'/4}
 \\[12pt]
 \chi\big[{\hat\R_{(\ell+1)p,p'}^{a,0}}\big](q)
   &=&\disp \frac{2}{\eta(q)}\sum_{k\in\mathbb{Z}}q^{(a+(2k+\ell)p)^2p'/4p}
 \\[12pt]
 \chi\big[{\hat\R_{p,(\ell+1)p'}^{0,b}}\big](q)
   &=&\disp \frac{2}{\eta(q)}\sum_{k\in\mathbb{Z}}q^{(b+(2k+\ell)p')^2p/4p'}
 \\[12pt]
 \chi[{\hat\R_{(\ell+1)p,p'}^{a,b}}](q)
  &=&\disp\frac{2}{\eta(q)}\sum_{k\in\mathbb{Z}}\big[q^{(ap'-bp+(2k+\ell)pp')^2/4pp'}
    \!+q^{(ap'+bp+(2k+\ell)pp')^2/4pp'}\big]
 \end{array}
\label{charProj}
\ee
where $\ell=0,1$, and they respect the identities
\be
 \begin{array}{rl}
 &\displaystyle{\chi\big[{\hat\R_{p,p'}^{a,0}}\big](q)= \chi\big[{\hat\R_{2p,p'}^{p-a,0}}\big](q),\qquad 
   \chi\big[{\hat\R_{p,p'}^{0,b}}\big](q)= \chi\big[{\hat\R_{p,2p'}^{0,p'-b}}\big](q)}
  \\[10pt]
 &\displaystyle{\chi[{\hat\R_{(3-\kappa)p,p'}^{a,b}}](q)=\chi[{\hat\R_{\kappa p,p'}^{p-a,b}}](q)
     =\chi[{\hat\R_{\kappa p,p'}^{a,p'-b}}](q)}
 \end{array}
\label{KacIden}
\ee 
The number of linearly independent ${\cal W}$-projective characters is thus given by $\tfrac{1}{2}(p+1)(p'+1)$.

%%%%%%%
\begin{figure}[p]
\psset{unit=.9cm}
%\raisebox{2.8cm}{$\mbox{\bf Dense Polymers/}\atop\mbox{\bf Symp Fermions:\ \,}$}\quad
\begin{center}
\begin{pspicture}(-1,-.3)(3,3.25)
 \psframe[linewidth=0pt,fillstyle=solid,fillcolor=lightlightblue](0,0)(2,3)
 \psframe[linewidth=0pt,fillstyle=solid,fillcolor=midblue](0,0)(2,1)
 \psframe[linewidth=0pt,fillstyle=solid,fillcolor=midblue](0,2)(2,3)
 \psgrid[gridlabels=0pt,subgriddiv=1](0,0)(2,3)
  \rput(.5,2.5){$\frac{3}{8}$}
 \rput(1.5,2.5){$-\frac{1}{8}$}
 \rput(.5,1.5){$0$}
 \rput(1.5,1.5){$0$}
 \rput(.5,.5){$-\frac{1}{8}$}
 \rput(1.5,.5){$\frac{3}{8}$}
{\color{blue}
 \rput(.5,-.5){$0$}
 \rput(1.5,-.5){$1$}
 \rput(2.5,-.5){$r$}
 \rput(-.5,.5){$0$}
 \rput(-.5,1.5){$1$}
 \rput(-.5,2.5){$2$}
  \rput(-.5,3.25){$s$}}
\end{pspicture}\hspace{-20pt}
%\begin{pspicture}(-1,-.3)(3,3.25)
% \psframe[linewidth=0pt,fillstyle=solid,fillcolor=lightlightblue](0,0)(2,3)
% \rput[bl](0,0){\psframe[linewidth=0pt,fillstyle=solid,fillcolor=lightpurple](0,0)(1,1)}
%%\rput[bl](2,0){\psframe[linewidth=0pt,fillstyle=solid,fillcolor=lightpurple](0,0)(1,1)}
%\rput[bl](1,1){\psframe[linewidth=0pt,fillstyle=solid,fillcolor=darkpurple](0,0)(1,1)}
%\rput[bl](0,2){\psframe[linewidth=0pt,fillstyle=solid,fillcolor=lightpurple](0,0)(1,1)}
% \psgrid[gridlabels=0pt,subgriddiv=1](0,0)(2,3)
%\psline[linewidth=1.2pt,arrowsize=8pt]{<-}(.5,.5)(1.5,1.5)
%\psline[linewidth=1.2pt,arrowsize=8pt]{<-}(.7,.7)(1.5,1.5)
%\psline[linewidth=1.2pt,arrowsize=8pt]{<-}(.5,2.5)(1.5,1.5)
%\psline[linewidth=1.2pt,arrowsize=8pt]{<-}(.7,2.3)(1.5,1.5)
%\psline[linewidth=1.2pt,arrowsize=8pt](.5,-.5)(1.5,-.5)
%\multiput(0,0)(1,0){2}{\psline[linewidth=1.2pt](.5,-.65)(.5,-.35)}
%\psline[linewidth=1.2pt,arrowsize=8pt]{<->}(-.5,.5)(-.5,2.5)
%\psline[linewidth=1.2pt,arrowsize=8pt]{<->}(-.5,.8)(-.5,2.2)
%\multiput(0,0)(0,1){3}{\psline[linewidth=1.2pt](-.65,.5)(-.35,.5)}
%\rput[r](.5,.5){\small $0$}
%\rput[l](1.55,1.5){\small $1$}
%\rput[r](.45,2.5){\small $2$}
%\end{pspicture}
\begin{pspicture}(-1,-.3)(3,3.25)
 \psframe[linewidth=0pt,fillstyle=solid,fillcolor=lightlightblue](0,0)(2,3)
 \rput[bl](0,0){\psframe[linewidth=0pt,fillstyle=solid,fillcolor=lightpurple](0,0)(1,1)}
\rput[bl](1,1){\psframe[linewidth=0pt,fillstyle=solid,fillcolor=darkpurple](0,0)(1,1)}
\rput[bl](0,2){\psframe[linewidth=0pt,fillstyle=solid,fillcolor=lightpurple](0,0)(1,1)}
 \psgrid[gridlabels=0pt,subgriddiv=1](0,0)(2,3)
 \psline[linewidth=1.2pt,arrowsize=8pt]{<-}(.5,.5)(.51,.51)
\psline[linewidth=1.2pt,arrowsize=8pt]{<-}(.65,.65)(.66,.66)
\psline[linewidth=1.pt,arrowsize=8pt](.8,.85)(1.45,1.5)
\psline[linewidth=1.pt,arrowsize=8pt](.8,.75)(1.55,1.5)
\psline[linewidth=1.2pt,arrowsize=8pt]{<-}(.5,2.5)(.6,2.4)
\psline[linewidth=1.2pt,arrowsize=8pt]{<-}(.65,2.35)(.7,2.3)
\psline[linewidth=1.pt,arrowsize=8pt](.8,2.25)(1.55,1.5)
\psline[linewidth=1.pt,arrowsize=8pt](.8,2.15)(1.45,1.5)
\psline[linewidth=1.pt,arrowsize=8pt](.5,-.55)(1.5,-.55)
\psline[linewidth=1.pt,arrowsize=8pt](.5,-.45)(1.5,-.45)
\multiput(0,0)(1,0){2}{\psline[linewidth=1.2pt](.5,-.65)(.5,-.35)}
\psline[linewidth=1.2pt,arrowsize=8pt]{<->}(-.5,.5)(-.5,2.5)
\psline[linewidth=1.2pt,arrowsize=8pt]{<->}(-.5,.8)(-.5,2.2)
\multiput(0,0)(0,1){3}{\psline[linewidth=1.2pt](-.65,.5)(-.35,.5)}
\rput[r](.4,.5){\small $0$}
\rput[l](1.55,1.5){\small $1$}
\rput[r](.45,2.5){\small $2$}
\end{pspicture}
\begin{pspicture}(-1,-.3)(3,4.25)
\psframe[linewidth=0pt,fillstyle=solid,fillcolor=lightlightblue](0,0)(3,4)
 \psframe[linewidth=0pt,fillstyle=solid,fillcolor=lightestblue](1,1)(2,3)
 \multirput(0,0)(0,3){2}{\multirput(0,0)(2,0){2}{\psframe[linewidth=0pt,fillstyle=solid,fillcolor=midblue](0,0)(1,1)}}
 \psgrid[gridlabels=0pt,subgriddiv=1](0,0)(3,4)
 \rput(.5,3.5){$\frac{35}{24}$}
 \rput(1.5,3.5){$\frac{1}{3}$}
 \rput(2.5,3.5){$-\frac{1}{24}$}
  \rput(.5,2.5){$\frac{5}{8}$}
 \rput(1.5,2.5){$0$}
 \rput(2.5,2.5){$\frac{1}{8}$}
 \rput(.5,1.5){$\frac{1}{8}$}
 \rput(1.5,1.5){$0$}
  \rput(2.5,1.5){$\frac{5}{8}$}
 \rput(.5,.5){$-\frac{1}{24}$}
 \rput(1.5,.5){$\frac{1}{3}$}
  \rput(2.5,.5){$\frac{35}{24}$}
{\color{blue}
 \rput(.5,-.5){$0$}
 \rput(1.5,-.5){$1$}
 \rput(2.5,-.5){$2$}
 \rput(3.5,-.5){$r$}
 \rput(-.5,.5){$0$}
 \rput(-.5,1.5){$1$}
 \rput(-.5,2.5){$2$}
  \rput(-.5,3.5){$3$}
  \rput(-.5,4.25){$s$}}
\end{pspicture}\ \ 
\begin{pspicture}(-1,-.3)(3,4.25)
\psframe[linewidth=0pt,fillstyle=solid,fillcolor=lightlightblue](0,0)(3,4)
\rput[bl](0,0){\psframe[linewidth=0pt,fillstyle=solid,fillcolor=lightpurple](0,0)(1,1)}
\rput[bl](2,0){\psframe[linewidth=0pt,fillstyle=solid,fillcolor=lightpurple](0,0)(1,1)}
\rput[bl](1,1){\psframe[linewidth=0pt,fillstyle=solid,fillcolor=darkpurple](0,0)(1,1)}
\rput[bl](0,2){\psframe[linewidth=0pt,fillstyle=solid,fillcolor=lightpurple](0,0)(1,1)}
\rput[bl](2,2){\psframe[linewidth=0pt,fillstyle=solid,fillcolor=lightpurple](0,0)(1,1)}
\rput[bl](1,3){\psframe[linewidth=0pt,fillstyle=solid,fillcolor=darkpurple](0,0)(1,1)}
\psgrid[gridlabels=0pt,subgriddiv=1](0,0)(3,4)
\psline[linewidth=1.2pt,arrowsize=8pt]{<-}(.35,.35)(1.5,1.5)
\psline[linewidth=1.2pt,arrowsize=8pt]{<-}(.5,.5)(1.5,1.5)
\psline[linewidth=1.2pt,arrowsize=8pt]{<-}(.65,.65)(1.5,1.5)
\psline[linewidth=1.2pt,arrowsize=8pt]{<-}(.8,.8)(1.5,1.5)
\psline[linewidth=1.2pt,arrowsize=8pt]{<-}(.5,2.5)(1.5,1.5)
\psline[linewidth=1.2pt,arrowsize=8pt]{<-}(.7,2.3)(1.5,1.5)
\psline[linewidth=1.2pt,arrowsize=8pt]{<-}(2.65,.35)(1.5,1.5)
\psline[linewidth=1.2pt,arrowsize=8pt]{<-}(2.5,.5)(1.5,1.5)
\psline[linewidth=1.2pt,arrowsize=8pt]{<-}(2.35,.65)(1.5,1.5)
\psline[linewidth=1.2pt,arrowsize=8pt]{<-}(2.2,.8)(1.5,1.5)
\psline[linewidth=1.2pt,arrowsize=8pt]{<-}(2.5,2.5)(1.5,1.5)
\psline[linewidth=1.2pt,arrowsize=8pt]{<-}(2.3,2.3)(1.5,1.5)
\psline[linewidth=1.2pt,arrowsize=8pt](.5,2.55)(1.5,3.55)
\psline[linewidth=1.2pt,arrowsize=8pt](.5,2.45)(1.5,3.45)
\psline[linewidth=1.2pt,arrowsize=8pt](2.5,2.55)(1.5,3.55)
\psline[linewidth=1.2pt,arrowsize=8pt](2.5,2.45)(1.5,3.45)
\psline[linewidth=1.2pt,arrowsize=8pt]{<->}(.5,-.5)(2.5,-.5)
\psline[linewidth=1.2pt,arrowsize=8pt]{<->}(.75,-.5)(2.25,-.5)
\multiput(0,0)(1,0){3}{\psline[linewidth=1.2pt](.5,-.65)(.5,-.35)}
\psline[linewidth=1.2pt,arrowsize=8pt]{<->}(-.5,.5)(-.5,3.5)
\psline[linewidth=1.2pt,arrowsize=8pt]{<->}(-.5,.8)(-.5,3.2)
\multiput(0,0)(0,1){4}{\psline[linewidth=1.2pt](-.65,.5)(-.35,.5)}
\rput[r](.4,.6){\small $0$}
\rput[b](1.2,1.4){\small $1$}
\rput[b](1.8,1.4){\small $5$}
\rput[r](.4,2.5){\small $4$}
\rput[b](1.5,3.6){\small $3$}
\rput[l](2.6,2.5){\small $2$}
\rput[l](2.6,.6){\small $6$}
\end{pspicture}\\[20pt]
\end{center}
%%%%%%%%%%%%%%%%%%%%%%%%%%%%%%%%%%%%%%%%%%%%%%%%
%\mbox{}\hspace{.15in}
\begin{center}
\begin{pspicture}(-1,-.3)(6,8.25)
\psframe[linewidth=0pt,fillstyle=solid,fillcolor=lightlightblue](0,0)(5,8)
\psframe[linewidth=0pt,fillstyle=solid,fillcolor=lightestblue](1,1)(4,7)
\multirput(0,0)(0,7){2}{\multirput(0,0)(4,0){2}{\psframe[linewidth=0pt,fillstyle=solid,fillcolor=midblue](0,0)(1,1)}}
%\multirput(0,0)(0,2){4}{\multirput[bl](0,0)(2,0){3}{\psframe[linewidth=0pt,fillstyle=solid,fillcolor=lightpurple](0,0)(1,1)}}
%\multirput[bl](1,1)(0,2){4}{\psframe[linewidth=0pt,fillstyle=solid,fillcolor=darkpurple](0,0)(1,1)}
%\multirput[bl](3,1)(0,2){4}{\psframe[linewidth=0pt,fillstyle=solid,fillcolor=darkpurple](0,0)(1,1)}
\psgrid[gridlabels=0pt,subgriddiv=1](0,0)(5,8)
\rput(.5,.5){$-\frac{9}{112}$}
\rput(1.5,.5){$\frac{5}{14}$}
\rput(2.5,.5){$\frac{187}{112}$}
\rput(3.5,.5){$\frac{27}{7}$}
\rput(4.5,.5){$\frac{775}{112}$}%
\rput(.5,1.5){$\frac{1}{16}$}
\rput(1.5,1.5){$0$}
\rput(2.5,1.5){$\frac{13}{16}$}
\rput(3.5,1.5){$\frac{5}{2}$}
\rput(4.5,1.5){$\frac{81}{16}$}%
\rput(.5,2.5){$\frac{55}{112}$}
\rput(1.5,2.5){$-\frac{1}{14}$}
\rput(2.5,2.5){$\frac{27}{112}$}
\rput(3.5,2.5){$\frac{10}{7}$}
\rput(4.5,2.5){$\frac{391}{112}$}%
\rput(.5,3.5){$\frac{135}{112}$}
\rput(1.5,3.5){$\frac{1}{7}$}
\rput(2.5,3.5){$-\frac{5}{112}$}
\rput(3.5,3.5){$\frac{9}{14}$}
\rput(4.5,3.5){$\frac{247}{112}$}%
\rput(.5,4.5){$\frac{247}{112}$}
\rput(1.5,4.5){$\frac{9}{14}$}
\rput(2.5,4.5){$-\frac{5}{112}$}
\rput(3.5,4.5){$\frac{1}{7}$}
\rput(4.5,4.5){$\frac{135}{112}$}%
\rput(.5,5.5){$\frac{391}{112}$}
\rput(1.5,5.5){$\frac{10}{7}$}
\rput(2.5,5.5){$\frac{27}{112}$}
\rput(3.5,5.5){$-\frac{1}{14}$}
\rput(4.5,5.5){$\frac{55}{112}$}%
\rput(.5,6.5){$\frac{81}{16}$}
\rput(1.5,6.5){$\frac{5}{2}$}
\rput(2.5,6.5){$\frac{13}{16}$}
\rput(3.5,6.5){$0$}
\rput(4.5,6.5){$\frac{1}{16}$}%
\rput(.5,7.5){$\frac{775}{112}$}
\rput(1.5,7.5){$\frac{27}{7}$}
\rput(2.5,7.5){$\frac{187}{112}$}
\rput(3.5,7.5){$\frac{5}{14}$}
\rput(4.5,7.5){$-\frac{9}{112}$}%
{\color{blue}
 \rput(.5,-.5){$0$}
 \rput(1.5,-.5){$1$}
 \rput(2.5,-.5){$2$}
 \rput(3.5,-.5){$3$}
 \rput(4.5,-.5){$4$}
 \rput(5.5,-.5){$r$}
 \rput(-.5,.5){$0$}
 \rput(-.5,1.5){$1$}
 \rput(-.5,2.5){$2$}
  \rput(-.5,3.5){$3$}
   \rput(-.5,4.5){$4$}
 \rput(-.5,5.5){$5$}
 \rput(-.5,6.5){$6$}
  \rput(-.5,7.5){$7$}
  \rput(-.5,8.25){$s$}}
\end{pspicture}\hspace{-.25in}
\begin{pspicture}(-1,-.3)(6,8.25)
\psframe[linewidth=0pt,fillstyle=solid,fillcolor=lightlightblue](0,0)(5,8)
\multirput(0,0)(0,2){4}{\multirput[bl](0,0)(2,0){3}{\psframe[linewidth=0pt,fillstyle=solid,fillcolor=lightpurple](0,0)(1,1)}}
\multirput[bl](1,1)(0,2){4}{\psframe[linewidth=0pt,fillstyle=solid,fillcolor=darkpurple](0,0)(1,1)}
\multirput[bl](3,1)(0,2){4}{\psframe[linewidth=0pt,fillstyle=solid,fillcolor=darkpurple](0,0)(1,1)}
 \psgrid[gridlabels=0pt,subgriddiv=1](0,0)(5,8)
\psline[linewidth=1.2pt,arrowsize=8pt]{<->}(2.5,.5)(4.5,2.5)
\psline[linewidth=1.2pt,arrowsize=8pt]{<->}(2.7,.7)(4.3,2.3)
\psline[linewidth=1.2pt,arrowsize=8pt]{<->}(.35,.35)(4.5,4.5)
\psline[linewidth=1.2pt,arrowsize=8pt]{<->}(.5,.5)(4.3,4.3)
\psline[linewidth=1.2pt,arrowsize=8pt]{<->}(.65,.65)(4.5,4.5)
\psline[linewidth=1.2pt,arrowsize=8pt]{<->}(.8,.8)(4.3,4.3)
\psline[linewidth=1.2pt,arrowsize=8pt]{<->}(.5,2.5)(4.5,6.5)
\psline[linewidth=1.2pt,arrowsize=8pt]{<->}(.7,2.7)(4.3,6.3)
\psline[linewidth=1.2pt,arrowsize=8pt]{<->}(.5,4.5)(3.5,7.5)
\psline[linewidth=1.2pt,arrowsize=8pt]{<->}(.7,4.7)(3.3,7.3)
\psline[linewidth=1.2pt,arrowsize=8pt]{<->}(.5,2.5)(2.5,.5)
\psline[linewidth=1.2pt,arrowsize=8pt]{<->}(.7,2.3)(2.3,.7)
\psline[linewidth=1.2pt,arrowsize=8pt]{<->}(.5,4.5)(4.65,.35)
\psline[linewidth=1.2pt,arrowsize=8pt]{<->}(.7,4.3)(4.5,.5)
\psline[linewidth=1.2pt,arrowsize=8pt]{<->}(.5,4.5)(4.35,.65)
\psline[linewidth=1.2pt,arrowsize=8pt]{<->}(.7,4.3)(4.2,.8)
\psline[linewidth=1.2pt,arrowsize=8pt]{<->}(.5,6.5)(4.5,2.5)
\psline[linewidth=1.2pt,arrowsize=8pt]{<->}(.7,6.3)(4.3,2.7)
\psline[linewidth=1.2pt,arrowsize=8pt]{<->}(1.5,7.5)(4.5,4.5)
\psline[linewidth=1.2pt,arrowsize=8pt]{<->}(1.7,7.3)(4.3,4.7)
\psline[linewidth=1.2pt,arrowsize=8pt](.5,6.55)(1.5,7.55)
\psline[linewidth=1.2pt,arrowsize=8pt](.5,6.45)(1.5,7.45)
\psline[linewidth=1.2pt,arrowsize=8pt](4.5,6.55)(3.5,7.55)
\psline[linewidth=1.2pt,arrowsize=8pt](4.5,6.45)(3.5,7.45)
\psline[linewidth=1.2pt,arrowsize=8pt]{<->}(.5,-.5)(4.5,-.5)
\psline[linewidth=1.2pt,arrowsize=8pt]{<->}(.8,-.5)(4.2,-.5)
\psline[linewidth=1.2pt,arrowsize=8pt]{<->}(-.5,.5)(-.5,7.5)
\psline[linewidth=1.2pt,arrowsize=8pt]{<->}(-.5,.8)(-.5,7.2)
\multiput(0,0)(1,0){5}{\psline[linewidth=1.2pt](.5,-.65)(.5,-.35)}
\multiput(0,0)(0,1){8}{\psline[linewidth=1.2pt](-.65,.5)(-.35,.5)}
\rput[r](.3,.5){\small $0$}
\rput[r](2.4,.4){\small $14$}
\rput[r](4.4,.4){\small $28$}
\rput[r](1.33,1.5){\small $3$}
\rput[l](1.63,1.5){\small $11$}
\rput[r](3.33,1.5){\small $17$}
\rput[l](3.63,1.5){\small $25$}
\rput[r](1.33,3.5){\small $5$}
\rput[l](1.63,3.5){\small $19$}
\rput[r](3.33,3.5){\small $9$}
\rput[l](3.63,3.5){\small $23$}
\rput[r](1.33,5.5){\small $13$}
\rput[l](1.63,5.5){\small $27$}
\rput[r](3.33,5.5){\small $1$}
\rput[l](3.63,5.5){\small $15$}
\rput[b](1.5,7.6){\small $21$}
\rput[b](3.5,7.6){\small $7$}
\rput[r](.45,2.5){\small $8$}
\rput[r](2.33,2.5){\small $6$}
\rput[l](2.63,2.5){\small $22$}
\rput[l](4.55,2.5){\small $20$}
\rput[r](.45,4.5){\small $16$}
\rput[r](2.33,4.5){\small $2$}
\rput[l](2.63,4.5){\small $26$}
\rput[l](4.55,4.5){\small $12$}
\rput[r](.45,6.5){\small $24$}
\rput[r](2.33,6.5){\small $10$}
\rput[l](2.63,6.5){\small $18$}
\rput[l](4.55,6.5){\small $4$}
\end{pspicture}\\[20pt]
\end{center}
%%%%%%%%%%%%%%%%%%%%%%%%%
\begin{center}
\begin{pspicture}(-1,-.3)(5,6.25)
\psframe[linewidth=0pt,fillstyle=solid,fillcolor=lightlightblue](0,0)(4,6)
\multirput(0,0)(0,2){3}{\multirput[bl](0,0)(2,0){2}{\psframe[linewidth=0pt,fillstyle=solid,fillcolor=lightpurple](0,0)(1,1)}}
\multirput[bl](1,1)(2,0){2}{\psframe[linewidth=0pt,fillstyle=solid,fillcolor=darkpurple](0,0)(1,1)}
\multirput[bl](1,1)(2,0){2}{\psframe[linewidth=0pt,fillstyle=solid,fillcolor=darkpurple](0,0)(1,1)}
\multirput[bl](1,3)(2,0){2}{\psframe[linewidth=0pt,fillstyle=solid,fillcolor=darkpurple](0,0)(1,1)}
\multirput[bl](1,5)(2,0){2}{\psframe[linewidth=0pt,fillstyle=solid,fillcolor=darkpurple](0,0)(1,1)}
\psgrid[gridlabels=0pt,subgriddiv=1](0,0)(4,6)
\psline[linewidth=1.2pt,arrowsize=8pt]{<->}(.35,.35)(3.5,3.5)
\psline[linewidth=1.2pt,arrowsize=8pt]{<->}(.5,.5)(3.3,3.3)
\psline[linewidth=1.2pt,arrowsize=8pt]{<-}(.65,.65)(1.5,1.5)
\psline[linewidth=1.2pt,arrowsize=8pt]{<-}(.8,.8)(1.5,1.5)
\psline[linewidth=1.2pt,arrowsize=8pt]{<->}(.5,2.5)(3.35,5.35)
\psline[linewidth=1.2pt,arrowsize=8pt]{<->}(.7,2.7)(3.5,5.5)
\psline[linewidth=1.2pt,arrowsize=8pt]{<->}(.5,2.5)(3.65,5.65)
\psline[linewidth=1.2pt,arrowsize=8pt]{<->}(.7,2.7)(3.8,5.8)
\psline[linewidth=1.2pt,arrowsize=8pt](2.5,.55)(3.5,1.55)
\psline[linewidth=1.2pt,arrowsize=8pt](2.5,.45)(3.5,1.45)
\psline[linewidth=1.2pt,arrowsize=8pt](.5,4.55)(1.5,5.55)
\psline[linewidth=1.2pt,arrowsize=8pt](.5,4.45)(1.5,5.45)
\psline[linewidth=1.2pt,arrowsize=8pt]{<-}(.5,2.5)(1.5,1.5)
\psline[linewidth=1.2pt,arrowsize=8pt]{<-}(.7,2.3)(1.5,1.5)
\psline[linewidth=1.2pt,arrowsize=8pt]{<-}(2.3,.7)(1.5,1.5)
\psline[linewidth=1.2pt,arrowsize=8pt]{<-}(2.5,.5)(1.5,1.5)
\psline[linewidth=1.2pt,arrowsize=8pt]{<->}(.5,4.5)(3.5,1.5)
\psline[linewidth=1.2pt,arrowsize=8pt]{<->}(.7,4.3)(3.3,1.7)
\psline[linewidth=1.2pt,arrowsize=8pt]{<->}(1.5,5.5)(3.5,3.5)
\psline[linewidth=1.2pt,arrowsize=8pt]{<->}(1.7,5.3)(3.3,3.7)
\psline[linewidth=1.2pt,arrowsize=8pt]{<->}(.5,-.5)(3.5,-.5)
\psline[linewidth=1.2pt,arrowsize=8pt]{<->}(.75,-.5)(3.25,-.5)
\multiput(0,0)(1,0){4}{\psline[linewidth=1.2pt](.5,-.65)(.5,-.35)}
\psline[linewidth=1.2pt,arrowsize=8pt]{<->}(-.5,.5)(-.5,5.5)
\psline[linewidth=1.2pt,arrowsize=8pt]{<->}(-.5,.8)(-.5,5.2)
\multiput(0,0)(0,1){6}{\psline[linewidth=1.2pt](-.65,.5)(-.35,.5)}
\rput[r](.3,.5){\small $0$}
\rput[r](2.4,.4){\small $10$}
\rput[r](1.33,1.5){\small $2$}
\rput[l](1.63,1.5){\small $8$}
\rput[l](3.55,1.5){\small $12$}
\rput[r](1.33,3.5){\small $4$}
\rput[l](1.63,3.5){\small $14$}
\rput[l](3.63,3.5){\small $6$}
\rput[b](1.5,5.6){\small $10$}
\rput[b](3.3,5.6){\small $0$}
\rput[r](.45,2.5){\small $6$}
\rput[r](2.33,2.5){\small $4$}
\rput[l](2.63,2.5){\small $14$}
\rput[r](.45,4.5){\small $12$}
\rput[r](2.33,4.5){\small $2$}
\rput[l](2.63,4.5){\small $8$}
\end{pspicture}\hspace{-.25in}
%%%%%%%%%%%%%%%%%%%%%%%%%
\begin{pspicture}(-1,-.3)(5,6.25)
\psframe[linewidth=0pt,fillstyle=solid,fillcolor=lightlightblue](0,0)(4,6)
\multirput(0,0)(0,2){3}{\multirput[bl](0,0)(2,0){2}{\psframe[linewidth=0pt,fillstyle=solid,fillcolor=lightpurple](0,0)(1,1)}}
\multirput(1,1)(0,2){3}{\multirput[bl](0,0)(2,0){2}{\psframe[linewidth=0pt,fillstyle=solid,fillcolor=lightpurple](0,0)(1,1)}}
\multirput(1,0)(0,2){3}{\multirput[bl](0,0)(2,0){2}{\psframe[linewidth=0pt,fillstyle=solid,fillcolor=midblue](0,0)(1,1)}}
\psgrid[gridlabels=0pt,subgriddiv=1](0,0)(4,6)
\psline[linewidth=1.2pt,arrowsize=8pt,linecolor=blue]{<->}(.5,3.5)(2.5,5.5)
\psline[linewidth=1.2pt,arrowsize=8pt,linecolor=blue]{<->}(.7,3.7)(2.3,5.3)
\psline[linewidth=1.2pt,arrowsize=8pt,linecolor=blue]{<->}(.5,1.5)(3.5,4.5)
\psline[linewidth=1.2pt,arrowsize=8pt,linecolor=blue]{<->}(.7,1.7)(3.3,4.3)
\psline[linewidth=1.2pt,arrowsize=8pt,linecolor=blue]{<->}(1.5,.5)(3.5,2.5)
\psline[linewidth=1.2pt,arrowsize=8pt,linecolor=blue]{<->}(1.7,.7)(3.3,2.3)
\psline[linewidth=1.2pt,arrowsize=8pt,linecolor=blue](2.5,5.55)(3.5,4.55)
\psline[linewidth=1.2pt,arrowsize=8pt,linecolor=blue](2.5,5.45)(3.5,4.45)
\psline[linewidth=1.2pt,arrowsize=8pt,linecolor=blue](.5,1.55)(1.5,.55)
\psline[linewidth=1.2pt,arrowsize=8pt,linecolor=blue](.5,1.45)(1.5,.45)
\psline[linewidth=1.2pt,arrowsize=8pt,linecolor=blue]{<->}(.2,5.8)(3.5,2.5)
\psline[linewidth=1.2pt,arrowsize=8pt,linecolor=blue]{<->}(.35,5.65)(3.3,2.7)
\psline[linewidth=1.2pt,arrowsize=8pt,linecolor=blue]{<->}(.5,5.5)(3.5,2.5)
\psline[linewidth=1.2pt,arrowsize=8pt,linecolor=blue]{<->}(.65,5.35)(3.3,2.7)
\psline[linewidth=1.2pt,arrowsize=8pt,linecolor=blue]{<->}(.5,3.5)(3.65,.35)
\psline[linewidth=1.2pt,arrowsize=8pt,linecolor=blue]{<->}(.7,3.3)(3.5,.5)
\psline[linewidth=1.2pt,arrowsize=8pt,linecolor=blue]{<->}(.5,3.5)(3.35,.65)
\psline[linewidth=1.2pt,arrowsize=8pt,linecolor=blue]{<->}(.7,3.3)(3.2,.8)
\psline[linewidth=1.2pt,arrowsize=8pt]{<->}(.5,-.5)(3.5,-.5)
\psline[linewidth=1.2pt,arrowsize=8pt]{<->}(.75,-.5)(3.25,-.5)
\multiput(0,0)(1,0){4}{\psline[linewidth=1.2pt](.5,-.65)(.5,-.35)}
\psline[linewidth=1.2pt,arrowsize=8pt]{<->}(-.5,.5)(-.5,5.5)
\psline[linewidth=1.2pt,arrowsize=8pt]{<->}(-.5,.8)(-.5,5.2)
\multiput(0,0)(0,1){6}{\psline[linewidth=1.2pt](-.65,.5)(-.35,.5)}
\rput[r](1.3,.5){\small $5$}
\rput[r](3.4,.4){\small $15$}
\rput[r](.4,1.5){\small $3$}
\rput[r](2.33,1.5){\small $7$}
\rput[l](2.63,1.5){\small $13$}
\rput[r](1.33,2.5){\small $1$}
\rput[l](1.63,2.5){\small $11$}
\rput[l](3.6,2.5){\small $9$}
\rput[r](.44,3.5){\small $9$}
\rput[r](2.33,3.5){\small $1$}
\rput[l](2.63,3.5){\small $11$}
\rput[r](1.33,4.5){\small $7$}
\rput[l](1.63,4.5){\small $13$}
\rput[l](3.6,4.5){\small $3$}
\rput[b](.65,5.65){\small $15$}
\rput[b](2.5,5.65){\small $5$}
\end{pspicture}
\hspace{-.25in}
\begin{pspicture}(-1,-.3)(5,6.25)
\psframe[linewidth=0pt,fillstyle=solid,fillcolor=lightlightblue](0,0)(4,6)
\multirput[bl](1,3)(2,0){2}{\psframe[linewidth=0pt,fillstyle=solid,fillcolor=lightpurple](0,0)(1,1)}
\multirput[bl](0,4)(2,0){2}{\psframe[linewidth=0pt,fillstyle=solid,fillcolor=lightpurple](0,0)(1,1)}
\multirput[bl](1,5)(2,0){2}{\psframe[linewidth=0pt,fillstyle=solid,fillcolor=lightpurple](0,0)(1,1)}
\multirput[bl](0,0)(2,0){2}{\psframe[linewidth=0pt,fillstyle=solid,fillcolor=darkpurple](0,0)(1,1)}
\multirput[bl](1,1)(2,0){2}{\psframe[linewidth=0pt,fillstyle=solid,fillcolor=darkpurple](0,0)(1,1)}
\multirput[bl](0,2)(2,0){2}{\psframe[linewidth=0pt,fillstyle=solid,fillcolor=darkpurple](0,0)(1,1)}
\multirput[bl](0,3)(2,0){2}{\psframe[linewidth=0pt,fillstyle=solid,fillcolor=midblue](0,0)(1,1)}
\multirput[bl](1,4)(2,0){2}{\psframe[linewidth=0pt,fillstyle=solid,fillcolor= midblue](0,0)(1,1)}
\multirput[bl](0,5)(2,0){2}{\psframe[linewidth=0pt,fillstyle=solid,fillcolor= midblue](0,0)(1,1)}
\psgrid[gridlabels=0pt,subgriddiv=1](0,0)(4,6)
\psline[linewidth=1.2pt,arrowsize=8pt]{<-}(.35,.35)(2.5,2.5)
\psline[linewidth=1.2pt,arrowsize=8pt]{<-}(.5,.5)(2.5,2.5)
\psline[linewidth=1.2pt,arrowsize=8pt]{<-}(.65,.65)(1.5,1.5)
\psline[linewidth=1.2pt,arrowsize=8pt]{<-}(.8,.8)(1.5,1.5)
\psline[linewidth=1.2pt,arrowsize=8pt]{<-}(.5,2.5)(2.5,2.5)
\psline[linewidth=1.2pt,arrowsize=8pt]{<-}(.7,2.5)(2.5,2.5)
\psline[linewidth=1.2pt,arrowsize=8pt](2.5,.55)(3.5,1.55)
\psline[linewidth=1.2pt,arrowsize=8pt](2.5,.45)(3.5,1.45)
\psline[linewidth=1.2pt,arrowsize=8pt]{<-}(.5,2.5)(1.5,1.5)
\psline[linewidth=1.2pt,arrowsize=8pt]{<-}(.7,2.3)(1.5,1.5)
\psline[linewidth=1.2pt,arrowsize=8pt]{<-}(2.3,.7)(1.5,1.5)
\psline[linewidth=1.2pt,arrowsize=8pt]{<-}(2.5,.5)(1.5,1.5)
\psline[linewidth=1.2pt,arrowsize=8pt]{->}(2.5,2.5)(3.5,1.5)
\psline[linewidth=1.2pt,arrowsize=8pt,linecolor=blue]{<->}(.5,3.5)(2.5,5.5)
\psline[linewidth=1.2pt,arrowsize=8pt,linecolor=blue]{<->}(.7,3.7)(2.3,5.3)
\psline[linewidth=1.2pt,arrowsize=8pt,linecolor=blue]{->}(2.5,3.5)(3.5,4.5)
\psline[linewidth=1.2pt,arrowsize=8pt,linecolor=blue]{->}(2.5,3.5)(3.3,4.3)
\psline[linewidth=1.2pt,arrowsize=8pt,linecolor=blue](2.5,5.55)(3.5,4.55)
\psline[linewidth=1.2pt,arrowsize=8pt,linecolor=blue](2.5,5.45)(3.5,4.45)
\psline[linewidth=1.2pt,arrowsize=8pt,linecolor=blue]{<-}(.2,5.8)(2.5,3.5)
\psline[linewidth=1.2pt,arrowsize=8pt,linecolor=blue]{<-}(.35,5.65)(2.5,3.5)
\psline[linewidth=1.2pt,arrowsize=8pt,linecolor=blue]{<-}(.5,5.5)(2.5,3.5)
\psline[linewidth=1.2pt,arrowsize=8pt,linecolor=blue]{<-}(.65,5.35)(2.5,3.5)
\psline[linewidth=1.2pt,arrowsize=8pt,linecolor=blue]{<-}(.5,3.5)(2.5,3.5)
\psline[linewidth=1.2pt,arrowsize=8pt,linecolor=blue]{<-}(.7,3.5)(2.5,3.5)
\psline[linewidth=1.2pt,arrowsize=8pt]{->}(2.5,2.5)(3.3,1.7)
\psline[linewidth=1.2pt,arrowsize=8pt]{<->}(.5,-.5)(3.5,-.5)
\psline[linewidth=1.2pt,arrowsize=8pt]{<->}(.75,-.5)(3.25,-.5)
\multiput(0,0)(1,0){4}{\psline[linewidth=1.2pt](.5,-.65)(.5,-.35)}
\psline[linewidth=1.2pt,arrowsize=8pt]{<->}(-.5,.5)(-.5,5.5)
\psline[linewidth=1.2pt,arrowsize=8pt]{<->}(-.5,.8)(-.5,5.2)
\multiput(0,0)(0,1){6}{\psline[linewidth=1.2pt](-.65,.5)(-.35,.5)}
\pscircle[linewidth=1.2pt](2.5,2.72){.25}
\pscircle[linewidth=1.2pt,linecolor=blue](2.5,3.28){.25}
\rput[r](.3,.5){\small $0$}
\rput[r](2.4,.4){\small $10$}
\rput[r](1.33,1.5){\small $2$}
\rput[l](1.63,1.5){\small $8$}
\rput[l](3.55,1.5){\small $12$}
\rput[r](.45,2.5){\small $6$}
\rput[r](2.33,2.5){\small $4$}
\rput[l](2.63,2.5){\small $14$}
\rput[r](.44,3.5){\small $9$}
\rput[r](2.33,3.5){\small $1$}
\rput[l](2.63,3.5){\small $11$}
\rput[r](1.33,4.5){\small $7$}
\rput[l](1.63,4.5){\small $13$}
\rput[l](3.6,4.5){\small $3$}
\rput[b](.65,5.65){\small $15$}
\rput[b](2.5,5.65){\small $5$}
\end{pspicture}\\
\end{center}
\caption{Kac tables of conformal weights for the projective Grothendieck representations of symplectic fermions 
${\cal WLM}(1,2)$, critical percolation 
${\cal WLM}(2,3)$ and ${\cal WLM}(4,7)$. The corners, edges and interior entries are indicated by shading. 
The coset graphs $A^{(2)}_{p,p'}$, including ${\cal WLM}(3,5)$, are also shown where the loops are single loops. A pair of parallel lines indicates a double bond in both directions. 
For ${\cal WLM}(1,2)$, there are four arrows pointing towards each end node.
The pairs of integers at each node of a coset graph are the indices of $\varkappa^n_{rp'-sp}(q)$ and 
$\varkappa^n_{rp'+sp}(q)$. These indices agree at the corners and along the edges of each table. If $p+p'$ is odd, the 
${\mathbb Z}_2$ quotient gives a single coset graph on the $r+s$ even sublattice. If $p+p'$ is even, the ${\mathbb Z}_2$ 
quotient gives a pair of disconnected but isomorphic graphs sitting on separate sublattices as shown in the 
${\cal WLM}(3,5)$ case. 
\label{GrothKac}}
\end{figure}

In this context, a ring structure can be formed by moving to equivalence classes called projective Grothendieck generators where two ${\cal W}$-projective representations belong to the same equivalence class if and only if they share a common character. Specifically,  the character identities (\ref{KacIden}) are elevated to equivalence relations between the corresponding representations.  
Recalling  (\ref{Rp}), we thus define the $\tfrac{1}{2}(p+1)(p'+1)$ projective Grothendieck generators by
\be
 \Gc_{r,s}=[\Rch_{p,p'}^{r,s}],
  \qquad \Gc_{r,s}=\Gc_{p-r,p'-s},
  \qquad 0\le r \le p\ ,\quad 0\le s \le p'
\label{G}
\ee
To a projective Grothendieck generator, we assign the common character of the representatives within its equivalence class
\be
 \chit[\Gc_{r,s}](q)=\chit[\Rch_{p,p'}^{r,s}](q)
\ee
An equivalence class is uniquely characterized by the conformal weight
\be
 \Delta_{r,s}=\Delta_{p-r,p'-s}=\frac{(p'r-ps)^2-(p-p')^2}{4pp'},
   \quad\qquad 0\le r \le p\ ,\quad 0\le s \le p'
\label{DrsProj}
\ee
It follows that the projective Grothendieck generators can be organized into a Kac table
satisfying the ${\mathbb Z}_2$ Kac-table symmetry (\ref{Kacsym}). 
As shown in Figure~\ref{GrothKac}, the interior of this enlarged Kac table is the usual Kac table of the rational minimal 
model ${\cal M}(p,p')$ surrounded by a frame consisting of corners and edges. 
The equivalence classes of $\Wc$-projective rank-1 representations sit on 
the corners, rank-2 on the edges and rank-3 in the interior.
The projective Grothendieck Kac tables for symplectic fermions ${\cal WLM}(1,2)$, critical percolation ${\cal WLM}(2,3)$ 
and ${\cal WLM}(4,7)$ are shown 
as examples in Figure~\ref{GrothKac}. For $p=1$, the rational minimal Kac table is empty, there are no $\Wc$-projective rank-3  
representations and only the border remains. 

It follows from (\ref{KacIden}) that the degree (\ref{dimensions}) assigned to a projective 
Grothendieck generator $\Gc_{r,s}$ is simply the order of its equivalence class
\be
 d_{r,s}=\mbox{order}(\Gc_{r,s})=\big|[\Rch_{p,p'}^{r,s}]\big|,
    \qquad 0\le r \le p\ ,\quad 0\le s \le p'
 \label{dimG}
 \ee
 which means that
 \be
  \big|[\Pch_{\rh,\sh}]\big|= d_{\rh,\sh}=2^{\text{rank}(\Pch_{\rh,\sh})-1},
  \qquad 0\le \rh\le 2p-1,\quad 0\le \sh\le p'-1
\label{|P|}
\ee

Loosely speaking, the projective Grothendieck generators (\ref{G}) form a ring structure as 
the ``fusion algebra of characters". This ring is called the projective Grothendieck ring.
More precisely, the projective Grothendieck group is first defined as the free abelian group 
generated by the projective Grothendieck generators.
The group operation is addition and is defined via direct summation of the representations of
the equivalence classes
\be
 [\Pch_1]+[\Pch_2]=[\Pch_1\oplus \Pch_2]
\ee
that is, by addition of characters.
For rational CFTs, the Grothendieck group admits a ring structure whose multiplication 
follows from the fusion product of representations
$[R_1]\ast[R_2]=[R_1\otimes R_2]$.
For logarithmic models, on the other hand, the fusion of representations does not in general induce
a product on a Grothendieck group in this way, see \cite{GabRW09} for example.
However, on the {\em projective} Grothendieck group, the fusion rules {\em do} induce the multiplication
\be
 [\Pch_1]\ast[\Pch_2]=[\Pch_1\otimes \Pch_2]
\ee
turning the group into a ring. 

The explicit multiplication rules in the Grothendieck ring follow from the conjectured
fusion rules (\ref{FusProj}) and are given by
\be
 \Gc_{r,s}\ast\Gc_{r',s'}
  =d_{r,s}d_{r',s'}\!\!
   \sum_{r''=\eps(p+r+r'+1),\,\mathrm{by}\,2}^{p-\eps(r+r'+1)}\
   \sum_{s''=\eps(p'+s+s'+1),\,\mathrm{by}\,2}^{p'-\eps(s+s'+1)}\!\!
   \Gc_{r'',s''}
\label{GGG}
\ee
It follows that, up to the multiplicities $d_{r,s}d_{r',s'}\!\in\!\{1,2,4,8,16\}$, there are only {\em two} possible linear
combinations of generators arising as the result of a simple multiplication in the projective Grothendieck ring.
To appreciate this, we introduce the four generators
\be
 \Gamma^{\ell,\ell'}=\sum_{r=\eps(p-\ell),\,\mathrm{by}\,2}^{p-\ell}\
   \sum_{s=\eps(p'-\ell'),\,\mathrm{by}\,2}^{p'-\ell'}
   \Gc_{r,s},\qquad \ell,\ell'=0,1
\label{Gamma}
\ee
and note that
\be
 \begin{array}{llll}
 &\Gamma^{1,1}=\Gamma^{1,0}=\displaystyle{\sum_{r,s\ \mathrm{odd}}\Gc_{r,s}},\qquad
 &\Gamma^{0,1}=\Gamma^{0,0}=\displaystyle{\sum_{r,s\ \mathrm{even}}\Gc_{r,s}},\qquad 
 &p\ \mathrm{even},\ p'\ \mathrm{odd}
 \\[16pt]
 &\Gamma^{1,1}=\Gamma^{0,1}=\displaystyle{\sum_{r,s\ \mathrm{odd}}\Gc_{r,s}},\qquad
 &\Gamma^{1,0}=\Gamma^{0,0}=\displaystyle{\sum_{r,s\ \mathrm{even}}\Gc_{r,s}},\qquad
 &p\ \mathrm{odd},\ p'\ \mathrm{even}
 \\[16pt]
 &\Gamma^{1,1}=\Gamma^{0,0}=\displaystyle{\sum_{r+s\ \mathrm{even}\atop s\leq(p'-1)/2}\Gc_{r,s}},  
   \qquad
 &\Gamma^{0,1}=\Gamma^{1,0}=\displaystyle{\sum_{r+s\ \mathrm{odd}\atop s\leq(p'-1)/2}\Gc_{r,s}},\qquad
 &p,p'\ \mathrm{odd}
 \end{array}
\label{GammaGamma}
\ee
This observation will play an important role in Section~\ref{SecGrothBoundary}, as will the
character associated to $\Gamma^{\ell,\ell'}$
\be
 \chit[\Gamma^{\ell,\ell'}](q)=\sum_{r=\eps(p-\ell),\,\mathrm{by}\,2}^{p-\ell}\
   \sum_{s=\eps(p'-\ell'),\,\mathrm{by}\,2}^{p'-\ell'}
   \chit[\Gc_{r,s}](q)
\label{chiGamma}
\ee

\section{Boundary CFT and Coset Graphs}
\label{SecBoundary}

\subsection{$c=1$ revisited}
\label{Secc1}

The affine $u(1)$ models with central charge $c=1$ have been well studied \cite{Ginsparg,Kiritsis,DVVV,CappelliApp}. 
In particular, the $c=1$ boson on the circle $S^1$, with radius of compactification 
$R=\sqrt{2p'/p}$ expressed in terms of the coprime integers $p,p'$, 
exhibits an extended symmetry with $2n=2pp'$ primary operators $\phi_j=\exp(ij\varphi/\sqrt{n})$. 
The conformal weights are
\be
 \Delta_j=\mbox{min}\Big[\frac{j^2}{4n},\frac{(2n-j)^2}{4n}\Big],\qquad j=0,1,\ldots,2n
\ee 
with the corresponding affine $u(1)$ characters given by
\be
 \varkappa^n_j(q)=\varkappa^n_{2n-j}(q)=\frac{\Theta_{j,n}(q)}{\eta(q)}
  =\frac{1}{\eta(q)} \sum_{k\in{\mathbb Z}} q^{(j+2kn)^2/4n}
\label{KM}
\ee
The modular $S$-matrix follows from the modular $S$-transformations
\be
 \varkappa^n_j(e^{-2\pi i/\tau})=\sum_{k=0}^{2n-1} S_{jk} \varkappa^n_k(e^{2\pi i\tau}) 
  =\frac{1}{\sqrt{2n}}\sum_{k=0}^{2n-1} e^{-\pi i jk/n} \varkappa^n_k(e^{2\pi i\tau})
\label{KMmod}
\ee
The modular invariant partition functions are
\be
 Z^{\text{Circ}}_{p,p'}(q)=\sum_{j=0}^{2n-1} 
   \varkappa^n_j(q)\varkappa^n_{\omega_0 j}(\qb)
\label{ZCirc}
\ee
where the Bezout number $\omega_0$ is defined by
\be
 \omega_0=r_0p'+s_0p\quad (\mbox{mod $2n$})
\ee
in terms of the Bezout pair $(r_0,s_0)$ which is uniquely determined by the conditions
\be
 r_0p'-s_0p=1,\qquad\quad 1\le r_0\le p-1,\quad 1\le s_0\le p'-1,\quad p s_0<p'r_0
\ee
Any integer $j$ can be written as $j=ap'+bp$ and satisfies
\be
 \omega_0(ap'+bp)=ap'-bp\quad (\mbox{mod $2n$})
\ee
It follows that multiplication of an index $j$ of an 
affine $u(1)$ character by the Bezout number $\omega_0$ is an involution satisfying 
\be
 \varkappa^n_{\omega_0(r p'\pm s p)}(q)=\varkappa^n_{r p'\mp s p}(q),\qquad 
    r=0,1,\ldots,p;\quad s=0,1,\ldots, p'
\ee
The modular invariant is diagonal if $p=1$ (in which case $\omega_0=1$). 
In terms of $p,p'$, the duality $R\to2/R$ amounts to invariance of the partition function
under $p\leftrightarrow p'$.

The fusion algebra is given by the ${\mathbb Z}_{2n}$ algebra 
\be
 \phi_i\times \phi_j=\sum_{k=0}^{2n-1} N_{ij}{}^k\phi_k,\qquad N_{ij}{}^k=\delta_{i+j,k}^{(2n)}
\ee
where $i,j,k$ and their sums are interpreted as integers mod $2n$. In accord with the $u(1)$ symmetry, representations of 
this algebra are given by powers of the cyclic shift matrix $\Omega^\omega$ where $\omega$ is coprime to 
$2n=2pp'$ and $\Omega^{2n}=I$.
This theory is associated with the cyclic directed graph ${\mathbb Z}_{2n}$ with $2n$ nodes. Alternatively, if for $j\ne 0$ 
we work with the operators $\phi_j+\phi_{-j}=2\cos(j\varphi/\sqrt{n})$, we obtain the affine Dynkin diagram $A^{(1)}_{2n}$ 
with $2n$ nodes as shown in Figure~\ref{Aintertwine}. Representations of this fusion algebra are given by $N_0=I$, 
$N_1=\Omega^\omega+\Omega^{-\omega}$ along with the $sl(2)$ recursion $N_j=N_{j-1}N_1-N_{j-2}$.
\begin{figure}[htbp]
\begin{center}
\begin{pspicture}[shift=-3.05](-1.5,.8)(4,6.5)
%\psgrid(4,6)
\pscircle(2,4){2}
\qdisk(0,4){.08}
\qdisk(4,4){.08}
\qdisk(2,2){.08}
\qdisk(2,6){.08}
\qdisk(3.414,5.414){.08}
\qdisk(3.414,2.595){.08}
\qdisk(.595,5.414){.08}
\qdisk(.595,2.595){.08}
\psline[linewidth=1.pt,arrowsize=8pt]{<->}(0,1)(4,1)
\psline[linewidth=1.pt,arrowsize=8pt]{<->}(.3,1)(3.7,1)
\multirput(0,1)(1,0){5}{\psline[linewidth=1.2pt](0,-.12)(0,.12)}
\rput[c](0,.6){\small $0$}
\rput[c](1,.6){\small $1$}
\rput[c](2,.6){\small $2$}
\rput[c](3,.6){\small $3$}
\rput[c](4,.6){\small $4$}
\rput[c](0,1.3){\color{blue}\small $1$}
\rput[c](1,1.3){\color{blue}\small $1,2$}
\rput[c](2,1.3){\color{blue}\small $1,2$}
\rput[c](3,1.3){\color{blue}\small $1,2$}
\rput[c](4,1.3){\color{blue}\small $1$}
\rput[r](-.1,4){\small $0$}
\rput[r](.495,5.595){\small $1$}
\rput[b](2,6.15){\small $2$}
\rput[l](3.514,5.595){\small $3$}
\rput[l](4.1,4){\small $4$}
\rput[l](3.514,2.595){\small $5$}
\rput[t](2,1.85){\small $6$}
\rput[r](.455,2.595){\small $7$}
\rput[l](.1,4){\color{blue}\small $1$}
\rput[l](.695,5.295){\color{blue}\small $1$}
\rput[t](2,5.85){\color{blue}\small $1$}
\rput[r](3.314,5.295){\color{blue}\small $1$}
\rput[r](3.9,4){\color{blue}\small $1$}
\rput[r](3.314,2.695){\color{blue}\small $1$}
\rput[b](2,2.15){\color{blue}\small $1$}
\rput[l](.75,2.695){\color{blue}\small $1$}
\rput[c](-1.5,4.05){$A^{(1)}_{2n}$}
\rput[c](-1.5,1.05){$A^{(2)}_n$}
\end{pspicture}
\qquad\qquad
\mbox{$C=\begin{pmatrix}
2&0&0&0&0\\
0&1&0&0&0\\
0&0&1&0&0\\
0&0&0&1&0\\
0&0&0&0&2\\
0&0&0&1&0\\
0&0&1&0&0\\
0&1&0&0&0\end{pmatrix}$}
\end{center}
\caption{The intertwining relation between the affine $A^{(1)}_{2n}$ and twisted affine $A^{(2)}_n$ Dynkin graphs 
illustrated for $n=4$. 
The adjacency matrices satisfy the intertwining relation $A^{(1)}_{2n} C=CA^{(2)}_n$. The entries of the left
and right Perron-Frobenius eigenvectors are also shown.
\label{Aintertwine}}
\end{figure}

\subsection{Twisted affine coset graphs}
\label{SecTwisted}

Boundary rational CFTs are classified by graphs~\cite{BPPZ1,BPPZ2}. 
Since the ${\cal WLM}(p,p')$ models resemble 
rational theories, it is natural to ask to what extent their properties are also encoded in graphs. 
Since the affine Lie algebra $A^{(1)}_{2n}$ is simply-laced, the adjacency matrix of its Dynkin 
diagram is symmetric. 
Accordingly, the associated lattice transfer matrices are normal and diagonalizable. But this is not the 
case for the logarithmic minimal models which possess transfer matrices which are not normal 
and are not diagonalizable. 
We assert (see Section~\ref{SecModularVerlinde}) 
that the relevant graph for ${\cal WLM}(1,p)$ is the non-simply-laced Dynkin diagram of the 
{\em twisted} affine Lie algebra $A^{(2)}_p$~\cite{Kac} as shown in Figure~\ref{Aintertwine}. 
If there is an intertwining relation~\cite{PZhou} between these graphs that intertwine the largest eigenvalues 
then the theories described by the graphs have a common effective central charge. 
Since the largest eigenvalues are intertwined here, the associated CFTs have the same effective
central charge $c^{\text{eff}}=1$. We further observe that this intertwining relation also intertwines 
exponents and their associated characters. 
The eigenvalues $\lambda^p_r$ of the $(p+1)$-dimensional
fundamental adjacency matrix $N_1$ of $A^{(2)}_p$ are 
\be
N_1=\begin{pmatrix}
0&1&&&&&\\
2&0&1&&&&\\
&1&\cdot&&&&\\[-6pt]
&&&\hspace{-6pt}\cdot\hspace{-6pt}&&&\\[-6pt]
&&&&\cdot&1&\\
&&&&1&0&2\\
&&&&&1&0\\
\end{pmatrix}, \qquad \lambda^p_r=2\cos \frac{r\pi}{p},\qquad r=0,1,\ldots,p
\ee
The transpose of this non-symmetric adjacency matrix $A=N_1^T$ is related to 
the {\it symmetrizable} (generalized) Cartan matrix \mbox{$C=2I-A$}. 
Therefore, there exists a diagonal matrix 
\be
 D=\mbox{diag}(d_0,d_1,\ldots, d_p)=\mbox{diag}(1,2,2,\ldots,2,1)%,\qquad d_r\in{\mathbb N}
\ee
such that $B=DA$ is symmetric. It follows that $D^{1/2}AD^{-1/2}=D^{-1/2}BD^{-1/2}$ so that $A$ and $A^T$ are similar to 
real symmetric matrices with real eigenvalues and eigenvectors. Let
\be
 \vec a =(1,2,2,\ldots,2,1),\qquad \check{\vec a}=(1,1,1,\ldots,1,1)
\ee
be the right and left Perron-Frobenius eigenvectors of $A$, respectively, with entries $a_r$ and $\check{a}_r$ given by the 
Coxeter and dual Coxeter exponents (labels) respectively. It is known~\cite{Kac} that for all $A$-$D$-$E$ symmetrizable twisted affine 
Cartan matrices
\be
 d_r=d_r^{(p)}=\frac{\check{a}_0}{a_0}\,\frac{a_r}{\check{a}_r}\in{\mathbb N},\qquad r=0,1,2,\ldots\quad
   \mbox{with\quad $d_0=1$}
\ee
For $A^{(2)}_p$, the dimensions $d_r=a_r$ coincide with the Coxeter exponents.

Introducing $X=N_1$, the $A_p^{(2)}$ graph algebra is generated by the $p+1$ matrices $N_r$
\be
 N_r=d_rT_r(\tfrac{X}{2}),\qquad 0\leq r\leq p
\ee
where $T_r(X)$ is the $r$'th Chebyshev polynomial of the first kind. Closure of the algebra is encoded
in the minimal polynomial
\be
 T_{p+1}(\tfrac{X}{2})-T_{p-1}(\tfrac{X}{2})=0
\ee
and the multiplication rules are given explicitly by
\be
 N_rN_{r'}=\sum_{r''=0}^p N_{rr'}{}^{r''}N_{r''}
  =\frac{d_rd_{r'}}{2}\Big(\frac{1}{d_{|r-r'|}}N_{|r-r'|}+\frac{1}{d_{p-|p-r-r'|}}N_{p-|p-r-r'|}\Big),\qquad
  0\leq r,r'\leq p
\label{NNrr}
\ee

Similarly, as shown in Figure~\ref{GrothKac} and discussed in Section~\ref{SecModularVerlinde}, 
we argue that the relevant graph for ${\cal WLM}(p,p')$ is the coset graph 
\be
 A^{(2)}_{p,p'}=A^{(2)}_p\otimes A^{(2)}_{p'}/{\mathbb Z}_2 
\label{Acoset}
\ee
where the ${\mathbb Z}_2$ quotient is taken with respect to the $\mathbb{Z}_2$ Kac-table 
symmetry. This graph has $\tfrac{1}{2}(p+1)(p'+1)$ nodes. 
%We make the identification $A^{(2)}_{1,p}\equiv A^{(2)}_p$. 
The eigenvalues of the coset graph adjacency matrices are
\be
 \lambda^{p,p'}_{r,s}=4\cos\frac{r\pi}{p}\cos\frac{s\pi}{p'},\qquad r=0,1,\ldots,p;\quad s=0,1,\ldots,p'
\ee
where $\lambda^{p,p'}_{r,s}=\lambda^{p,p'}_{p-r,p'-s}$ reflects the $\mathbb{Z}_2$ Kac-table symmetry. 
For the symmetrizable coset graphs $A^{(2)}_{p,p'}$, the dimensions are given by
\be
d_{r,s}=d_r^{(p)} d_s^{(p')}=
\begin{cases} 
1,& \mbox{$(r,s)$ is a corner}\\[-2pt]
2,& \mbox{$(r,s)$ is on an edge}\\[-2pt]
4, & \mbox{$(r,s)$ is in the interior}
\end{cases}
\ee
and agree with (\ref{dimensions}) and (\ref{dimG}). 
In this way, the dimensions of the $\mbox{Hom}$ spaces (\ref{HomSpaces}) and ranks of the projective covers (\ref{ranks}) 
are related (through the 
Coxeter exponents) to data of twisted affine Dynkin diagrams. 
We also observe that the dimension $d_{r,s}$ coincides with the coordination number of the node $(r,s)$ of the coset 
graph where the coordination number is defined as the number of distinct nearest neighbours including itself if there is a 
loop. 
In particular, the matrix $D$ for critical percolation ${\cal WLM}(2,3)$ is given by
\be
 D=\mathrm{diag}(d_{0,0},d_{1,1},d_{2,2},d_{1,3},d_{0,2},d_{2,0})=\mathrm{diag}(1,4,2,2,2,1)
\label{D23}
\ee

Introducing $X=N_{1,0}$ and $Y=N_{0,1}$, the $A_{p,p'}^{(2)}$ graph algebra 
\be
 N_{r,s} N_{r',s'} =\sum_{r'',s''}N_{rs,r's'}{}^{r''s''} N_{r'',s''}
\label{NNNN}
\ee
is generated by the $\tfrac{1}{2}(p+1)(p'+1)$ matrices $N_{r,s}=N_{p-r,p'-s}$
\be
 N_{r,s}=d_{r,s}T_r(\tfrac{X}{2})T_s(\tfrac{Y}{2}),\qquad 0\leq r\leq p;\quad 0\leq s\leq p'
\label{NTT}
\ee
where 
\be
 T_{p+1}(\tfrac{X}{2})-T_{p-1}(\tfrac{X}{2})
 =T_{p'+1}(\tfrac{Y}{2}\big)-T_{p'-1}(\tfrac{Y}{2})
 =T_p(\tfrac{X}{2})-T_{p'}(\tfrac{Y}{2})=0
\label{mod}
\ee
The last equation reflects the $\mathbb{Z}_2$ Kac-table symmetry. 
One verifies that the sets $\{N_{r,0};\ r=0,\ldots,p\}$ and
$\{N_{0,s};\ s=0,\ldots,p'\}$ provide representations of the $A_p^{(2)}$ and $A_{p'}^{(2)}$ graph algebras, respectively. 
The structure constants in (\ref{NNNN}) are given explicitly by
\be
 {N_{r s, r' s'}}^{r'' s''}=\frac{d_{r,s}d_{r',s'}}{4d_{r'',s''}}\big(\delta_{r'',|r-r'|}+\delta_{r'',p-|p-r-r'|}\big)
    \big(\delta_{s'',|s-s'|}+\delta_{s'',p'-|p'-s-s'|}\big)
\label{Nddd}
\ee

\subsection{$\Wc$-projective and minimal characters in terms of affine $u(1)$ characters}
\label{SecProjKacMoody}

It follows from (\ref{charProj}) that the $\Wc$-projective characters are simple combinations of $c^{\text{eff}}=1$ 
affine $u(1)$ characters 
\bea
 \chi\big[{\hat\R_{(\ell+1)p,p'}^{0,0}}\big](q)&=&\varkappa^n_{\ell n}(q)\nn
 \chi\big[{\hat\R_{(\ell+1)p,p'}^{a,0}}\big](q)&=&2\varkappa^n_{ap'+ \ell n}(q)\nn
 \chi\big[{\hat\R_{p,(\ell+1)p'}^{0,b}}\big](q)&=&2\varkappa^n_{bp+ \ell n}(q)\nn
 \chi[{\hat\R_{(\ell+1)p,p'}^{a,b}}](q)&=&4\times\half\big[\varkappa^n_{ap'-bp+ \ell n}(q)+\varkappa^n_{ap'+bp+\ell n}(q)\big]
\eea
where $n=pp'$, $\ell=0,1$, $a=1,2,\ldots, p-1$, $b=1,2,\ldots,p'-1$ and the indices of the 
affine $u(1)$ characters are interpreted mod $2n$. 
The projective Grothendieck Kac table is formed by taking into account the $\mathbb{Z}_2$ 
Kac-table symmetry and the identities (\ref{KacIden}).
To each position of these Kac tables (Figure~\ref{GrothKac}), we associate 
a {\em pair} of characters consisting of one $\Wc$-projective and one minimal character
\be
 \chi[{\Gc_{r,s}}](q)=d_{r,s} \varkappa^n_{r,s}(q),\qquad \ch_{r,s}(q)
\ee
(recalling (\ref{Rp})) where 
\be
\begin{array}{rcl}
 \varkappa^n_{r,s}(q)%=\varkappa^n_{\D_{r,s}}(q)
   &=&\half\big[\varkappa^n_{rp'-sp}(q)+\varkappa^n_{rp'+sp}(q)\big],
  \\[8pt]
 \ch_{r,s}(q)%=\ch_{\D_{r,s}}(q)
   &=&\varkappa^n_{rp'-sp}(q)-\varkappa^n_{rp'+sp}(q),\end{array}
   \quad\qquad 0\leq r\leq p;
    \quad 0\leq s\leq p'
\label{charKM}
\ee
and $\ch_{r,s}(q)$ is the usual minimal Virasoro character (\ref{ch}) if $(r,s)$ is in the interior and 
vanishes otherwise. 
It is also noted that $\varkappa^n_{rp'-sp}(q)=\varkappa^n_{rp'+sp}(q)$ on the corners 
and edges of the Kac table and that, as defined in the interior of the Kac table, 
$\varkappa^n_{r,s}(q)$ has half-integer coefficients as a $q$ series. This definition turns out to be 
useful, while all physical character expressions built from $\varkappa^n_{r,s}(q)$ will involve only 
integer $q$ series. 
We also note that the character of the projective cover $\Pc_{a,b}$
of a minimal $\Wc$-irreducible can be written as
\be
 \chit[\Pc_{a,b}](q)=3\varkappa^n_{ap'-bp}(q)-\varkappa^n_{ap'+bp}(q),\qquad
   1\leq a\leq p-1;    \quad 1\leq b\leq p'-1
\ee

The coset graphs $A^{(2)}_{p,p'}$ can be viewed as built from linear $A^{(2)}_n$-type graphs with 
$n+1=pp'+1$ nodes by gluing pairs of nodes together. 
One linear graph starts in the corner $(r,s)=(0,0)$ and proceeds straight up and to the right 
through $(r,s)=(1,1)$ until it encounters an edge at which point it reflects turning through 90 
degrees. This continues until the graph terminates at a corner after visiting all $\half(p+1)(p'+1)$ 
nodes on the $r+s$ even sublattice at least once. On the given sublattice, the corner and edge nodes are visited exactly 
once and the interior nodes are visited exactly twice. Pairs of nodes on the linear graph 
corresponding to crossings in the interior of the Kac table (caused by visiting the same node of the 
coset graph twice) are glued together. Along the graph, the pair of integers at each node are given 
by the indices of $\varkappa^n_{k(p'-p)}(q)$ and $\varkappa^n_{\omega_0 k(p'-p)}(q)$ with 
$k=0,1,\ldots, pp'$. These two integers agree on an edge or a corner in which case it is 
written only once. Another linear graph with $pp'+1$ nodes, starting at either
$(p,0)$ or $(0,p')$ on the $r+s$ odd sublattice, similarly fills up the odd sublattice. Along this 
graph, the pair of integers at each node are given by the indices of $\varkappa^n_{k(p'-p)+pp'}(q)$ 
and $\varkappa^n_{\omega_0(k(p'-p)+pp')}(q)$ with $k=0,1,\ldots, pp'$. In the case $p+p'$ even, 
nodes with the same pair of indices are identified in which case the coset graphs are formed by 
reflection restricted to the lower or upper halves of the Kac table as shown in 
Figure~\ref{GrothKac}. This interpretation of the coset graph introduces a natural linear order and 
labelling of the nodes by $j=0,1,\ldots,\half(p+1)(p'+1)-1$ according to the order in which the 
distinct nodes are visited. We will use this labelling extensively in this paper and already 
employed it in (\ref{D23}). With this labelling, $j=0$ labels the identity and $j=1$ labels the 
fundamental of the graph fusion algebra.

The building of the coset graph $A^{(2)}_{p,p'}$ from the linear graph $A^{(2)}_{pp'}$ can be formalized as 
an ``intertwining similarity" relation. Symbolically, this takes the form
\be
A^{(2)}_{p,p'}=A^{(2)}_p\otimes A^{(2)}_{p'}/{\mathbb Z}_2=2C_L A^{(2)}_{pp'} C_R\label{IntSim}
\ee
where the ``intertwining matrices" $C_L$ and $C_R$ are rectangular and not square. This relation is a 
similarity relation in the sense that $C_L$ is a generalized left inverse of $C_R$, or equivalently, $C_R$ 
is a generalized right inverse of $C_L$
\be
C_L C_R=I
\ee
The relation is an intertwining relation in the sense that the common eigenvalues of the coset graph $A^{(2)}_{p,p'}$ 
and the linear graph $2A^{(2)}_{pp'}$ are intertwined. Explicitly, the relation between the eigenvalues is given by the identity
\be
 \lambda^{p,p'}_{r,s}=4\cos\frac{r\pi}{p}\cos\frac{s\pi}{p'}
  =2\cos\frac{(rp'-sp)\pi}{pp'}+2\cos\frac{(rp'+sp)\pi}{pp'}
  =\lambda^{pp'}_{rp'-sp} + \lambda^{pp}_{rp'+sp}
\label{eigvalId}
\ee
%where $k$, $r$ and $s$ are related by
%\be
%\varkappa^n_{rp'-sp}(q)=\varkappa^n_{k(p'-p)}(q)
%\ee
The points in the interior of the 
Kac table (Figure~\ref{GrothKac}), where the linear graph crosses itself to form the coset graph, are labelled 
by pairs of distinct indices. The common eigenvalues are those which occur on the frame of the Kac table where the indices 
in the pair coincide
%, $\varkappa^n_{k(p'-p)}(q)=\varkappa^n_{\omega_0k(p'-p)}(q)$, 
and the two terms on the right side of (\ref{eigvalId}) agree. 
The action of the intertwining similarity (\ref{IntSim}) is to implement a change of basis 
to symmetric and anti-symmetric combinations of these pairs of indices and to project out the 
anti-symmetric combinations (\ref{charKM}) according to
\be
pp'+1-\half(p-1)(p'-1)=\half(p+1)(p'+1)
\ee
For this reason the anti-symmetric combinations, corresponding to the minimal 
characters $\ch_{r,s}(q)$, are like twisted sectors. They are not relevant in the boundary theory, since the corresponding representations by themselves do not have boundary conditions associated with them, but they do appear 
in the modular invariant partition function of the bulk theory. For $p+p'$ odd, a typical example is
${\cal WLM}(2,3)$. In this case, we have explicitly
\bea
A^{(2)}_{pp'}=\mbox{\scriptsize{\mbox{$\bordermatrix{
&0&1&2&3&4&5&6\cr
0&0&1&0&0&0&0&0\cr
1&2&0&1&0&0&0&0\cr
2&0&1&0&1&0&0&0\cr
3&0&0&1&0&1&0&0\cr
4&0&0&0&1&0&1&0\cr
5&0&0&0&0&1&0&2\cr
6&0&0&0&0&0&1&0}$}}},\qquad\ 
A^{(2)}_{p,p'}=\mbox{\scriptsize{\mbox{$\bordermatrix{
&0&1&2&3&4&6\cr
0&0&1&0&0&0&0\cr
1&4&0&2&0&2&4\cr
2&0&1&0&2&0&0\cr
3&0&0&2&0&2&0\cr
4&0&1&0&2&0&0\cr
6&0&1&0&0&0&0}$}}}\\
C_L=\mbox{\scriptsize{\mbox{$\bordermatrix{
&0&1&2&3&4&5&6\cr
0&1&0&0&0&0&0&0\cr
1&0&1&0&0&0&1&0\cr
2&0&0&1&0&0&0&0\cr
3&0&0&0&1&0&0&0\cr
4&0&0&0&0&1&0&0\cr
6&0&0&0&0&0&0&1}$}}},\qquad\ \ 
2C_R=\mbox{\scriptsize{\mbox{$\bordermatrix{
&0&1&2&3&4&6\cr
0&2&0&0&0&0&0\cr
1&0&1&0&0&0&0\cr
2&0&0&2&0&0&0\cr
3&0&0&0&2&0&0\cr
4&0&0&0&0&2&0\cr
5&0&1&0&0&0&0\cr
6&0&0&0&0&0&2}$}}}
\eea
For $p+p'$ even, a typical example is
${\cal WLM}(3,5)$ and some modifications are needed. 
As seen in Figure~\ref{GrothKac}, the linear and coset graphs break up into direct sums of two identical graphs
\be
A^{(2)}_{pp'}\mapsto T^{(2)}_{(pp'-1)/2}\oplus T^{(2)}_{(pp'-1)/2},\qquad A^{(2)}_{p,p'}=G\oplus G
\ee
where the tadpole $T^{(2)}_{(pp'-1)/2}$ is obtained from the folding ($\mathbb{Z}_2$ orbifolding) of $A^{(2)}_{pp'}$ 
\be
\begin{pspicture}[shift=-0.25](-1.5,.6)(4,1.5)
\pscircle[linewidth=1pt](4,1.25){.25}
\psline[linewidth=1.pt,arrowsize=8pt]{<-}(0,1)(4,1)
\psline[linewidth=1.pt,arrowsize=8pt]{<-}(.3,1)(3.7,1)
\multirput(0,1)(1,0){5}{\psline[linewidth=1.2pt](0,-.12)(0,.12)}
\rput[c](0,.6){\small $0$}
\rput[c](1,.6){\small $1$}
\rput[c](2,.6){\small $2$}
\rput[c](3.05,.6){\small $\cdots$}
\rput[c](4,.6){\small $n$}
\rput[c](-1.5,1.05){$T^{(2)}_n$:}
\end{pspicture}
\ee
and $G\oplus G$ is the disconnected pair of graphs in the bottom-right of Figure~\ref{GrothKac}. 
In this case, focussing on the first block, we have explicitly
\bea
T^{(2)}_{(pp'-1)/2}=\mbox{\scriptsize{\mbox{$\bordermatrix{
&0&2&4&6&8&10&12&14\cr
\ 0&0&1&0&0&0&0&0&0\cr
\ 2&2&0&1&0&0&0&0&0\cr
\ 4&0&1&0&1&0&0&0&0\cr
\ 6&0&0&1&0&1&0&0&0\cr
\ 8&0&0&0&1&0&1&0&0\cr
10&0&0&0&0&1&0&1&0\cr
12&0&0&0&0&0&1&0&1\cr
14&0&0&0&0&0&0&1&1}$}}},\qquad\qquad
G=\mbox{\scriptsize{\mbox{$\bordermatrix{
&0&2&4&6&10&12\cr
\ 0&0&1&0&0&0&0\cr
\ 2&4&0&1&2&2&0\cr
\ 4&0&1&1&2&0&2\cr
\ 6&0&1&1&0&0&0\cr
10&0&1&0&0&0&2\cr
12&0&0&1&0&2&0}$}}}\\
C_L=\mbox{\scriptsize{\mbox{$\bordermatrix{
&0&2&4&6&8&10&12&14\cr
\ 0&1&0&0&0&0&0&0&0\cr
\ 2&0&1&0&0&1&0&0&0\cr
\ 4&0&0&1&0&0&0&0&1\cr
\ 6&0&0&0&1&0&0&0&0\cr
10&0&0&0&0&0&1&0&0\cr
12&0&0&0&0&0&0&1&0}$}}},\qquad\ \ 
2C_R=\mbox{\scriptsize{\mbox{$\bordermatrix{
&0&2&4&6&10&12\cr
\ 0&2&0&0&0&0&0\cr
\ 2&0&1&0&0&0&0\cr
\ 4&0&0&1&0&0&0\cr
\ 6&0&0&0&2&0&0\cr
\ 8&0&1&0&0&0&0\cr
10&0&0&0&0&2&0\cr
12&0&0&0&0&0&2\cr
14&0&0&1&0&0&0}$}}}
\eea
Similarly, it is possible to project onto the anti-symmetric combinations to obtain the minimal coset graph $A_{p,p'}$ 
but with $C_L$ and $C_R$ having some negative entries.

\subsection {Modular transformations and Verlinde algebra}
\label{SecModularVerlinde}

It was shown in \cite{FGST06} that the minimal and $\Wc$-projective characters separately carry representations of the modular 
group. 
The modular matrix for the $\Wc$-projective characters is determined by the identities (\ref{charKM}) and the affine $u(1)$  modular 
transformation (\ref{KMmod}). Here we find that there exists an associated Verlinde algebra \cite{Ver88} and that it is identical to
the coset $A_{p,p'}^{(2)}$ graph algebra.

First, the modular $S$-matrix of the $\Wc$-projective characters of ${\cal WLM}(1,p)$ is given by
\be
 S^p_{rr'}=\frac{d_r}{\sqrt{2p}}\cos\frac{rr'\pi}{p},\qquad r,r'=0,1,\ldots,p
\label{modMatrix}
\ee
where $d_r=2-\delta_{r,0}^{(p)}$.
More generally, the modular matrix of the $\Wc$-projective characters of ${\cal WLM}(p,p')$ takes the coset form
\bea
 S^{p,p'}_{rs,r's'}
 &=&\sqrt{2}(-1)^{rs'+r's}S^p_{r,p'r'}S^p_{s,ps'}\nn
  &=&\frac{d_{r,s}}{\sqrt{2pp'}}(-1)^{rs'+r's}\cos\frac{rr'p'\pi}{p}\cos\frac{ss'p\pi}{p'}\\
  &=&\frac{d_{r,s}}{\sqrt{2pp'}}(-1)^{(r+s)(r'+s')}\cos\frac{rr'(p'-p)\pi}{p}\cos\frac{ss'(p'-p)\pi}{p'}\nonumber
\label{Spp}
\eea
where the pairs $j=(r,s)$ and $(r',s')$ are suitably restricted to representative pairs under the 
identification of the ${\mathbb Z}_2$ Kac-table symmetry. 
For cases with $p+p'$ odd, we restrict to the sublattice $r+s$ even. If $p+p'$ is even, we restrict 
to $r+s$ even for $s\le (p'-1)/2$ and $r+s$ odd for $s\ge (p'+1)/2$. This is merely a convenient 
choice as any equivalent restriction is allowed, but it has the advantage of being related to the 
labelling of the graphs embedded in the Kac tables in Figure \ref{GrothKac}.
The modular matrices satisfy $(S^{p,p'})^2=I$, $(S^{p,p'})^T\ne S^{p,p'}$, and it is noted
that $S^{1,p'}=S^{p'}$.

The modular matrix of the $\Wc$-projective characters gives rise to a Verlinde algebra 
\be
 N_i N_j =\sum_{k=0}^{\frac{1}{2}(p+1)(p'+1)-1} N_{ij}{}^k N_k
\label{VerlindeAlg}
\ee
where $i,j,k$ all run over the allowed pairs $(r,s)$, the fused adjacency matrices $N_i$ have entries 
$(N_i)_j{}^k=N_{ij}{}^k$ and the structure constants are
\be
 N_{ij}{}^k=\sum_{m=0}^{\frac{1}{2}(p+1)(p'+1)-1} \frac{S_{im}S_{jm}S_{mk}}{S_{0m}}\in {\mathbb N}_0
\label{structure}
\ee
where we have omitted the superscripts $p,p'$.
Since, $N_0=N_{0,0}=I$, the identity in this Verlinde algebra is $(r,s)=(0,0)$. 
To show that the algebra is indeed the $A^{(2)}_{p,p'}$ graph algebra
discussed in Section~\ref{SecTwisted}, one first observes that, by inserting the explicit expressions (\ref{Spp})
for the modular matrix entries, the structure constants
(\ref{structure}) separate into a product of two sums over triple products of cosines. Multiple applications
of the trigonometric identity
\be
 2\cos a\cos b=\cos(a+b)+\cos(a-b)
\ee
permits writing these sums as sums over single cosines.
These sums can be evaluated using the identity
\be
 \sum_{j=0}^{p}(2-\delta_{j,0}^{(p)})(-1)^{kj}\cos\frac{jm\pi}{p}=2p
 \Big(\delta_{m,0}^{(2p)}\eps(k+1)+\delta_{m,p}^{(2p)}\eps(k)\Big)
\ee
where $\eps(k)$ is given in (\ref{parity}).
A manipulation of the various (generalized) Kronecker deltas then shows
that the structure constants (\ref{structure}) of the Verlinde algebra
are identical to the structure constants (\ref{Nddd}) of the $A^{(2)}_{p,p'}$ graph algebra.

The fundamental matrix $N_1=N_{1,1}$
is the adjacency matrix of the $A^{(2)}_{p,p'}$ coset graph shown in Figure~\ref{GrothKac}.
The quantum dimensions 
\be
 \frac{S_{r s, m m'}}{S_{0 0, m m'}}=d_{r,s}(-1)^{rm'+sm}\cos\frac{rmp' \pi}{p}\cos\frac{sm'p\pi}{p'},
   \qquad 0\leq r,m\leq p;\quad 0\leq s,m'\leq p'
\ee
give a one-dimensional representation of the $A^{(2)}_{p,p'}$ Verlinde algebra.

\subsection{Grothendieck ring and boundary conformal partition functions}
\label{SecGrothBoundary}

Each of the ${\cal W}$-projective representations (\ref{R}) has an associated boundary condition
(identified using the disentangling procedure of \cite{Rasm0805}). 
The conformal partition functions on a strip, however, depend on spectra (characters) and are blind to the reducible yet 
indecomposable structures present in most of these representations. 
To describe the conformal partition functions associated with the boundary conditions, it thus suffices to work with the 
projective Grothendieck ring. It follows readily from Section~\ref{SecProjKacMoody} that
\be
 Z_{(r,s)|(r',s')}(q)=\chit[\Gc_{r,s}\ast\Gc_{r',s'}](q)
  =\!\!
   \sum_{r''=\eps(p+r+r'+1),\,\mathrm{by}\,2}^{p-\eps(r+r'+1)}\
   \sum_{s''=\eps(p'+s+s'+1),\,\mathrm{by}\,2}^{p'-\eps(s+s'+1)}\!\!
     d_{r,s}d_{r',s'}\chit[\Gc_{r'',s''}](q)
\label{Zrs}
\ee
where we recall that $\chit[\Gc_{r'',s''}](q)=d_{r'',s''} \varkappa_{r'',s''}(q)$.
We observe that the degrees $d_{r,s}$ also give a one-dimensional representation of the $A^{(2)}_{p,p'}$ Verlinde algebra
\be
 \sum_{r''=\eps(r+r')\,\mathrm{by}\,2}^{p-\eps(p+r+r')}\ \sum_{s''=\eps(s+s')\,\mathrm{by}\,2}^{p'-\eps(p'+s+s')}
     {N_{r s, r' s'}}^{r'' s''}d_{r'',s''}=d_{r,s}d_{r',s'}
\ee
Using this, we find
\be
 Z_{(r,s)|(r',s')}(q)=
    \sum_{r''=\eps(r+r')\,\mathrm{by}\,2}^{p-\eps(p+r+r')}\ \sum_{s''=\eps(s+s')\,\mathrm{by}\,2}^{p'-\eps(p'+s+s')}
     {N_{r s, r' s'}}^{r'' s''}d_{r'',s''}\chit[\Gamma^{\eps(r+r'+1),\eps(s+s'+1)}](q)
\label{sumGamma}
\ee
Also using
\be
 \begin{array}{rcll}
 \displaystyle{\sum_{r,s\ \mathrm{odd}}{N_{r s, r' s'}}^{r'' s''}}
  &=&d_{r',s'}\eps(r'+r'')\eps(s'+s''),
   \qquad  &p+p'\ \mathrm{odd}
 \\[14pt]
 \displaystyle{\sum_{\text{$r\!+\!s$ even}\atop\text{$s\le (p'\!-\!1)/2$}}{N_{r s, r' s'}}^{r'' s''}}
  &=&d_{r',s'}\eps(r'+r''+s'+s''+1)(1-\lfloor\tfrac{2s''}{p'+1}\rfloor),\qquad
  &p+p'\ \mathrm{even}
 \end{array}
\ee
the conformal partition function can thus be written as
\be
 Z_{i|j}(q)=\sum_{k=0}^{\frac{1}{2}(p+1)(p'+1)-1} N_{ij}{}^k (F\vec\chi[\Gc])_k(q)
\label{Zij}
\ee
where $i,j,k$ all run over the allowed pairs $(r,s)$, as described in Section~\ref{SecProjKac} and indicated
in Figure \ref{GrothKac}. 
The matrix
\be
F=\begin{cases}
\disp\mbox{}\ \ \sum_{\text{$r,s$ odd}} N_{r,s},\ &\mbox{$p+p'$ odd}\\[18pt]
\disp\sum_{\text{$r\!+\!s$ even}\atop\text{$s\le (p'\!-\!1)/2$}} N_{r,s},\ &\mbox{$p+p'$ even}
\end{cases}
\label{F}
\ee
acts on the column of characters $\vec\chit[\Gc]=\{\chit[\Gc_{r,s}]\}$ to form the two block characters
appearing in (\ref{sumGamma}), cf. (\ref{GammaGamma}). The rank of the matrix $F$ is therefore two. If $p+p'$ is even, $F$ breaks up into a direct sum of two identical matrices $F'$. 
For ${\cal WLM}(1,2)$, ${\cal WLM}(2,3)$ and ${\cal WLM}(3,5)$, respectively, we have
\bea
F=\begin{pmatrix}
0&1&0\\
2&0&2\\
0&1&0
\end{pmatrix};\quad 
F=\begin{pmatrix}
0&1&0&1&0&0\\
4&0&4&0&4&4\\
0&2&0&2&0&0\\
2&0&2&0&2&2\\
0&2&0&2&0&0\\
0&1&0&1&0&0
\end{pmatrix};\quad 
F=F'\oplus F',\quad F'=
\begin{pmatrix}
1&1&1&1&1&1\\
4&4&4&4&4&4\\
4&4&4&4&4&4\\
2&2&2&2&2&2\\
2&2&2&2&2&2\\
2&2&2&2&2&2
\end{pmatrix}
\eea
Explicitly,
\be
 (F\vec\chi[\Gc])_{r,s}(q)=\sum_{r''=\eps(p+r+1),\,\mathrm{by}\,2}^{p-\eps(r+1)}\
   \sum_{s''=\eps(p'+s+1),\,\mathrm{by}\,2}^{p'-\eps(s+1)}\!\!
     d_{r,s}\chit[\Gc_{r'',s''}](q)
\ee 

We have now demonstrated that the modular matrix (\ref{modMatrix}) diagonalizes the 
multiplication rules (\ref{GGG}) of the projective Grothendieck ring.
Verlinde-like formulas for logarithmic CFTs have been discussed before in the literature 
\cite{FHST0306,FK0705,GT0711,GabR08,PRR0907,Ras0908,Ras0911}.
However, unlike the one above based on $\Wc$-projective characters, the various results in these papers all involve
some generalization of the fundamental structure and link between modular data and fusion rules.

\section{Bulk CFT and $A$-Type Modular Invariants}
\label{SecBulk}

In a rational CFT, a modular invariant partition function can be written~\cite{CardyModInv} as a sesquilinear form 
in irreducible characters. 
Our basic assumption is that this carries over to the logarithmic minimal models, at least for the
invariants describing physical partition functions.
In the following, we thus assume that a modular invariant in ${\cal WLM}(p,p')$ can
be written as a sesquilinear form in ${\cal W}$-irreducible characters
\be
 Z=\sum_{i,j\in\text{Irr}}M_{ij}\chit_i(q)\chit_j(\qb)
\label{Zirr}
\ee
It is recalled that there are $2pp'+\tfrac{1}{2}(p-1)(p'-1)$ $\Wc$-irreducible characters, cf. (\ref{ch}) and (\ref{chih}).
In the next subsection, we consider the implications of (\ref{Zirr}) in terms of ${\cal W}$-projective and minimal ${\cal W}$-irreducible characters.

\subsection{Separation into $\Wc$-projective and minimal characters}
\label{SecSeparation}

{\bf Proposition}\ \ \ An $S$-invariant sesquilinear form in ${\cal W}$-irreducible characters
can be expressed as a sesquilinear form in ${\cal W}$-projective and minimal 
${\cal W}$-irreducible characters.
\\[.2cm]
{\bf Proof}\ \ \ See Appendix \ref{AppProp1}.
\\[.2cm]
Even without invoking invariance under $T$-transformations,
we have thus reduced the analysis of modular invariants from the
$2pp'+\tfrac{1}{2}(p-1)(p'-1)$ $\Wc$-irreducible characters to the $\tfrac{1}{2}(p+1)(p'+1)$ 
$\Wc$-projective and $\tfrac{1}{2}(p-1)(p'-1)$ minimal $\Wc$-irreducible characters, a total of 
$pp'+1$ linearly independent characters. We note that this is the same number as the number of 
linearly independent affine $u(1)$ characters associated with the $c=1$ boson compactified on a 
circle of radius $R=\sqrt{2p'/p}$ in Section~\ref{Secc1}.
At this point, no assumptions have been made about the usual integrality of the coefficients
in the sesquilinear forms.

The next conjecture suggests that a modular invariant sesquilinear form in 
${\cal W}$-projective and minimal ${\cal W}$-irreducible characters can be written in such a 
way that there is no coupling between the two types of characters.
\\ \goodbreak
\noindent{\bf Conjecture}\ \ \ A modular invariant sesquilinear form in ${\cal W}$-projective and 
minimal ${\cal W}$-irreducible characters can be written as a sum of a modular invariant 
sesquilinear form in ${\cal W}$-projective characters and a modular invariant sesquilinear form 
in minimal ${\cal W}$-irreducible characters
\be
 Z=Z^{\mathrm{Proj}}+Z^{\mathrm{Min}}
\label{ZZZ}
\ee

This conjecture is supported by the following evidence. For $p=1$, the rational Kac table is empty, hence $Z^{\mathrm{Min}}=0$, and the conjecture
is trivially true. For $p\geq2$, we have tested it for a variety of models and all with affirmative outcome.
The evidence is summarized in Table \ref{tableppp}.

Combining the proposition and the conjecture above, it follows that {\em every} modular invariant $Z$ (which can be written as a sesquilinear form in 
${\cal W}$-irreducible characters) can be written as a sum of a modular invariant $Z^{\mathrm{Proj}}$ in 
${\cal W}$-projective characters and a modular invariant $Z^{\mathrm{Min}}$
in minimal ${\cal W}$-irreducible characters with {\em no} coupling between the 
${\cal W}$-projective and the minimal ${\cal W}$-irreducible characters. 
As we will see below, in particular in (\ref{nonsymZ}),
alternative expressions exist and can be very insightful, but they can always be rewritten
as in (\ref{ZZZ}).

If $Z^{\mathrm{Proj}}=0$, we are left with a modular invariant for the rational minimal models.
Our interest is in the {\em logarithmic} minimal models, so we assume that
$Z^{\mathrm{Proj}}\neq0$. Guided by rational CFT, it is then natural to expect that 
the operator with minimal conformal weight (\ref{Dmin}), the effective vacuum, enters exactly 
once in the modular invariant. We will assume this here, but hope to return to this issue elsewhere.
\begin{table}[t]
\begin{center}
$
\renewcommand{\arraystretch}{1.2}
\begin{array}{|c||c|c|c|c|c|c|c|c|c|c|c|c|c|c|c|c|c|c|c|cl}
\hline
p&2&3&4&5&6&7&8&9&10&11&12&13&14
\\[1.5pt]
\hline
p'_{\max}&111&74&55&44&37&32&27&25&21&20&17&17&15
\\[1.5pt]
\hline
\end{array}
$
\end{center}
\caption{Evidence for the separation $Z=Z^{\mathrm{Proj}}+Z^{\mathrm{Min}}$. For the given 
values of $p$, the separation has been tested in the affirmative for all $p'$ coprime to $p$ and 
satisfying $p<p' \leq p'_{\max}$, that is, for all $pp'\leq 225$.}
\label{tableppp}
\end{table}

\subsection{$A$-type modular invariants}
\label{SecAType}

Again guided by the observed structures in rational CFT, 
we expect the same coset graphs to classify ${\cal WLM}(p,p')$ on the strip and on the torus. 
Indeed, the graphs considered above reproduce the diagonal $A$-type modular invariants 
in $\Wc$-projective characters considered in \cite{FGST06,Wood10}
\be
 Z^{\text{Proj}}_{p,p'}(q)=\half\sum_{r=0}^p\sum_{s=0}^{p'}d_{r,s}\,|\varkappa_{r,s}(q)|^2
  =\half\sum_{r=0}^p\sum_{s=0}^{p'}\frac{1}{d_{r,s}}\,\big|\chi[{\cal G}_{r,s}](q)\big|^2
  =\sum_{\rh=0}^{2p-1}\sum_{\sh=0}^{p'-1} \frac{1}{d_{\rh,\sh}^2}\,|\chi[\Pch_{\rh,\sh}](q)|^2
\label{ZProj}
\ee
Here and in the following, we suppress the dependence on the superscript $n=pp'$ of 
$\varkappa$.
The factors of $\tfrac{1}{2}$ in the first two double sums in (\ref{ZProj}) reflect the 
$\mathbb{Z}_2$ Kac-table symmetry, and it is noted that $|\varkappa_{0,0}(q)|^2$ appears with 
multiplicity 1. When expanded out in the form (\ref{Zirr}), all multiplicities are nonnegative 
integers.
The required modular invariance of $Z^{\text{Proj}}_{p,p'}(q)$ follows immediately after 
realizing that $Z^{\text{Proj}}_{p,p'}(q)$ can be expressed in terms of the modular invariant
partition functions (\ref{ZCirc}) for the compactified boson
\be
 Z^{\mathrm{Proj}}_{p,p'}(q)=\half\big[Z^{\mathrm{Circ}}_{1,pp'}(q)+Z^{\mathrm{Circ}}_{p,p'}(q)\big] 
\label{ZPC}
\ee
It follows, in particular, that $Z^{\mathrm{Proj}}_{1,p'}(q)=Z^{\mathrm{Circ}}_{1,p'}(q)$.

Likewise, the coset graphs encode the diagonal $A$-type modular invariants of the 
rational minimal models
\be
 Z^{\mathrm{Min}}_{p,p'}(q)=\half\sum_{r=1}^{p-1}\sum_{s=1}^{p'-1} |\mbox{ch}_{r,s}(q)|^2
  =\half\sum_{r=0}^{p}\sum_{s=0}^{p'} \big|\varkappa_{rp'-sp}(q)-\varkappa_{rp'+sp}(q)\big|^2
\label{ZMin}
\ee
where the factors of $\tfrac{1}{2}$ again reflect the corresponding $\mathbb{Z}_2$ 
Kac-table symmetry.
In terms of the partition functions (\ref{ZCirc}) for the compactified boson, 
these invariants can be written as
\be
 Z^{\mathrm{Min}}_{p,p'}(q)=\half\big[Z^{\mathrm{Circ}}_{1,pp'}(q)-Z^{\mathrm{Circ}}_{p,p'}(q)\big]
\label{ZMC}
\ee

Assuming that the operator with minimal conformal weight enters exactly once, 
the modular invariant partition function of ${\cal WLM}(p,p')$ must take the form 
\be
 Z_{p,p'}(q)=Z^{\text{Proj}}_{p,p'}(q)+n_{p,p'} Z^{\text{Min}}_{p,p'}(q),\qquad n_{p,p'}\in {\mathbb Z}
\label{Zpp}
\ee
According to Section~\ref{SecSeparation}, a similar linear combination always applies,
even if $Z^{\text{Proj}}_{p,p'}(q)$ and $Z^{\text{Min}}_{p,p'}(q)$ are not given by the $A$-type expressions 
(\ref{ZProj}) and (\ref{ZMin}). We note that $n_{1,p'}$ is arbitrary since the rational Kac table
is empty in that case and hence $Z^{\text{Min}}_{1,p'}(q)=0$.
\\[.2cm]
\noindent {\bf Conjecture}\ \ \ For the $A$-type modular invariant partition function
(\ref{Zpp}), the constant $n_{p,p'}$ is given by
\be
 n_{p,p'}=2
\label{npp}
\ee

Let us gather some evidence in favour of this conjecture.
First, it is precisely for $n_{2,3}=2$ that we recover the only well-justified result for $p>1$ found in the literature, 
namely the modular invariant partition function for ${\cal WLM}(2,3)$ obtained recently in \cite{GRW10}.

More generally, with the projective covers given in Section~\ref{SecProjCover}, we see that for $n_{p,p'}=2$
\be
 \frac{1}{d_{a,b}}|\chit[\Gc_{a,b}](q)|^2+n_{p,p'}|\ch_{a,b}(q)|^2=
   \sum_{k=1}^4\chit[\Wc(\D^k_{a,b})](q)\,\chit[\Rch_{p,p'}^{a,b}](\qb)
     +\ch_{a,b}(q)\,\chit[\Pc_{a,b}](\qb)
\ee
where $1\leq a\leq p-1$ and $1\leq b\leq p'-1$.
Using (\ref{Wprojreps}) as well, this implies that
\be
 Z_{p,p'}(q)=\sum_{i\in\text{Irr}} \chit_i(q)\,\chit[\Pc_i](\qb)
\label{nonsymZ}
\ee
where the sum is over all $\Wc$-irreducible 
representations and $\Pc_i$ denotes the projective cover of $i$.
This is in accordance with the assumption made in
\cite{GRW10} that the space of bulk states on the torus 
(for the diagonal $A$ case) respects
\be
 {\cal H}_{\text{torus}}=\mathop{\bigoplus}_{i\in\text{Irr}}{{\cal W}_i\otimes \overline{{\cal P}_i}}
\ee
as an $(L_0,\bar{L}_0)$-graded vector space.
It is stressed, though, that the participating nonchiral representations cannot in general be written 
as tensor products of irreducible representations and their projective covers. 
Although the partition function (\ref{nonsymZ}) appears left-right asymmetric,
it is left-right symmetric when expanded in affine $u(1)$ characters.

Finally, by combining (\ref{ZPC}) and (\ref{ZMC}), we see that
\be
 Z_{p,p'}(q)=Z_{1,pp'}^{\mathrm{Circ}}(q)+(n_{p,p'}-1)Z_{p,p'}^{\mathrm{Min}}(q)
\ee
This is in accordance with
\be
 Z_{1,pp'}^{\mathrm{Circ}}(q)=Z_{p,p'}^{\mathrm{Proj}}(q)+Z_{p,p'}^{\mathrm{Min}}(q)
\ee
which follows directly from
\be
 d_{a,b}|\varkappa_{a,b}(q)|^2+|\ch_{a,b}(q)|^2=2|\varkappa_{ap'-bp}(q)|^2
   +2|\varkappa_{ap'+bp}(q)|^2,\qquad 1\leq a\leq p-1;\quad 1\leq b\leq p'-1
\ee
Now, if $n_{p,p'}=1$, the partition function
$Z_{p,p'}$ is merely given by the partition function for a compactified boson on a circle
of radius $R=\sqrt{2pp'}$.
Our conjecture thus yields a `minimal extension' of the partition function for
the compactified boson by
adding the partition function for the rational {\em minimal\/} model with the {\em minimal\/} 
positive integer coefficient, namely $n_{p,p'}-1=1$.

We conclude this section by listing a few specific modular invariant partition functions
of the form (\ref{Zpp}) with $n_{p,p'}=2$.
We label the various characters by their conformal weights, 
$\varkappa_{\Delta_{r,s}}(q)=\varkappa_{r,s}(q)$ and $\ch_{\Delta_{r,s}}(q)=\ch_{r,s}(q)$,
and suppress the dependence on $n=pp'$.
For symplectic fermions ${\cal WLM}(1,2)$, critical
percolation ${\cal WLM}(2,3)$, the logarithmic Ising model ${\cal WLM}(3,4)$ and
${\cal WLM}(3,5)$, the modular invariant partition functions are
\bea
 Z_{1,2}(q)\!\!\!&=&\!\!\!|\varkappa_{-\frac{1}{8}}(q)|^2+2|\varkappa_0(q)|^2
   +|\varkappa_{\frac{3}{8}}(q)|^2
 \nn
 Z_{2,3}(q)\!\!\!&=&\!\!\!|\varkappa_{-\frac{1}{24}}(q)|^2+4|\varkappa_{0}(q)|^2
   +2|\varkappa_{\frac{1}{8}}(q)|^2+2|\varkappa_{\frac{1}{3}}(q)|^2+2|\varkappa_{\frac{5}{8}}(q)|^2
   +|\varkappa_{\frac{35}{24}}(q)|^2+2|\ch_0(q)|^2
 \nn
 Z_{3,4}(q)\!\!\!&=&\!\!\!  |\varkappa_{-\frac{1}{48}}(q)|^2+4|\varkappa_{0}(q)|^2
    +4|\varkappa_{\frac{1}{16}}(q)|^2+2|\varkappa_{\frac{1}{6}}(q)|^2
    +2|\varkappa_{\frac{5}{16}}(q)|^2+4|\varkappa_{\frac{1}{2}}(q)|^2
    +2|\varkappa_{\frac{35}{48}}(q)|^2
 \nn
  &&\!\!\! +\,  2|\varkappa_{\frac{21}{16}}(q)|^2+2|\varkappa_{\frac{5}{3}}(q)|^2
    +|\varkappa_{\frac{143}{48}}(q)|^2
    +2\,\big\{|\ch_0(q)|^2+|\ch_{\frac{1}{16}}(q)|^2+|\ch_{\frac{1}{2}}(q)|^2\big\}
 \\
 Z_{3,5}(q)\!\!\!&=&\!\!\!   |\varkappa_{-\frac{1}{15}}(q)|^2+4|\varkappa_{0}(q)|^2
    +4|\varkappa_{\frac{1}{5}}(q)|^2+2|\varkappa_{\frac{7}{3}}(q)|^2
    +2|\varkappa_{\frac{8}{5}}(q)|^2+2|\varkappa_{\frac{8}{15}}(q)|^2
 \nn
  &&\!\!\! +\,    |\varkappa_{-\frac{221}{60}}(q)|^2+4|\varkappa_{\frac{3}{4}}(q)|^2
    +4|\varkappa_{-\frac{1}{20}}(q)|^2+2|\varkappa_{\frac{1}{12}}(q)|^2
    +2|\varkappa_{\frac{7}{20}}(q)|^2+2|\varkappa_{\frac{77}{60}}(q)|^2
 \nn
  &&\!\!\! +\,   2\,\big\{|\ch_0(q)|^2+|\ch_{\frac{1}{5}}(q)|^2+|\ch_{\frac{3}{4}}(q)|^2
  +|\ch_{-\frac{1}{20}}(q)|^2\big\}
 \nonumber
\eea

\section{Discussion}
\label{SecDiscussion}

In this paper, we have argued that, for the logarithmic minimal model ${\cal WLM}(p,p')$, 
the ${\cal W}$-projective representations are fundamental building blocks in both the bulk and 
boundary variants of the theory. For the boundary theory, in contrast to the ${\cal W}$-irreducible 
representations, there are boundary conditions associated with each of the ${\cal W}$-projective 
representations \cite{Rasm0805} arising from fundamental fusion. 
Here the associated boundary conformal partition functions 
were obtained by moving to the Grothendieck ring. 
This led to compact Verlinde-like formulas for the boundary partition functions 
involving $A$-type twisted affine graphs $A^{(2)}_p$ and their coset graphs 
\mbox{$A^{(2)}_{p,p'}=A^{(2)}_p\otimes A^{(2)}_{p'}/{\mathbb Z}_2$}. We argued that these graphs 
are classifying graphs for these theories 
playing a similar role to the classifying graphs for rational theories. This makes more explicit the 
sense in which these logarithmic theories resemble rational theories.

Guided by observed structures in rational theories, it is natural to expect that precisely the same 
coset graphs should appear in the modular invariant partition functions. On the torus, we 
generalized the work of \cite{GRW10} to conjecture modular invariant partition functions for all 
diagonal $A$-type ${\cal WLM}(p,p')$ theories as sesquilinear forms in ${\cal W}$-projective and 
rational minimal characters.  Indeed, we observe that they are naturally encoded by precisely the 
same coset graphs with striking relations to $c^{\text{eff}}=1$ Gaussian theories.  

Although we have argued that the coset graphs $A^{(2)}_{p,p'}$ are relevant {\em classifying graphs} in the bulk and boundary logarithmic minimal models, we stress that these graphs are classifying in a more limited sense than applies to rational CFTs. 
Specifically, in the rational case, the nodes of the classifying graph are in one-to-one correspondence with the 
irreducible representations and their conjugate boundary conditions. In the logarithmic setting, the representation content is more diverse and the nodes of the classifying graph are only in correspondence with the equivalence classes of boundary conditions associated with the ${\cal W}$-projective representations. Notwithstanding this feature, the classifying graphs  of the logarithmic minimal models (at least for these $A$-type theories) seem to completely encode the modular invariant partition functions. At present, the set of all possible non-$A$ type classifying graphs is not known.

This paper holds out possibilities for a deeper understanding and even a general 
classification of logarithmic minimal models. 
But there are as many questions left unanswered in this paper as answered. 
First, there is the question as to whether twisted affine Lie algebras have a role to play. So far, the algebraic 
approach to logarithmic CFT has focussed on ${\cal W}$-algebra structures. 
Second, the appearance of coset graphs begs the question as to whether there is an algebraic 
construction of these logarithmic 
theories  mimicking the Goddard, Kent and Olive construction~\cite{GKO85,GKO86} 
of rational minimal coset theories.
Third, there is as yet no evidence from the lattice approach to support the modular invariant 
partition functions (\ref{Zpp}) and (\ref{npp}). 
Such a check would firmly establish that these theories are physical on the torus. 
Fourth, there is a classification of twisted affine Lie algebras based on Dynkin graphs. It is therefore natural 
to ask if there are solvable lattice models and CFTs related to the Dynkin graphs which realize this classification. 
Lastly, it would be of interest to investigate the algebra of integrable seams on the torus and the 
relation to an Ocneanu-type algebra or its Grothendieck ring. We hope to return to some of these 
problems elsewhere.

%%%%%%%%%%%%%%%%%%%%%%%%%%%%%%%%%%%%%%%%%%%%%%
%
\section*{Acknowledgments}
\vskip.1cm
\noindent
This work is supported by the Australian Research Council (ARC). 
The authors thank Jean-Bernard Zuber for discussions and a critical reading of the manuscript.

%%%%%%%%%%%%%%%%%%%%%%%%%%%%%%%%%%%%%%%%%%%%%

\appendix

\section{Evidence for Separation in Section~\ref{SecSeparation}}
\label{AppEvidence}

\subsection{Modular Transformations of $\Wc$-Irreducible Characters}
\label{AppModular}

Following \cite{FGST06}, we consider the $2pp'+\tfrac{1}{2}(p-1)(p'-1)$ 
$\Wc$-irreducible characters
\be
 \{\chit_{r,s}^+(\tau)=\chit_{2p-r,s}(\tau),\ \chit_{r,s}^-(\tau)=\chit_{3p-r,s}(\tau);\ 
   (r,s)\in\Jc\},\qquad \{\chit_{r,s}(\tau);\ (r,s)\in\Jc_1\}
\ee
where we have introduced
\bea
 \Jc&=&\{(r,s);\ 1\leq r\leq p,\ 1\leq s\leq p'\}\nn
 \Jc_1&=&\{(r,s);\ 1\leq r\leq p-1,\ 1\leq s\leq p'-1,\ rp'+ps\leq pp'\}
\eea
The corresponding conformal weights are
\be
 \D(\chit_{r,s}^+)=\D_{r,2p'-s}=\D_{2p-r,s},\qquad
 \D(\chit_{r,s}^-)=\D_{r,3p'-s}=\D_{3p-r,s}\qquad
 \D(\chit_{r,s})=\D_{r,s}
\ee
We are using the same notation for a character as a function of $\tau$ as we are
for the same character as a function of $q$, but hope that this will not cause confusion.
The full set of $\Wc$-irreducible characters does not form a representation of the modular group,
whereas the minimal Virasoro characters $\ch_{r,s}(\tau)$ do.
To express the modular transformations of the $\Wc$-irreducible characters in a compact way, 
we introduce
\be
 \rtt=\frac{\pi p'rr'}{p},\qquad \stt=\frac{\pi pss'}{p'}
\ee
Under the modular $S$-transformation $\tau\to-\tfrac{1}{\tau}$, the characters transform as
\bea
 \chit_{r,s}^+(-\tfrac{1}{\tau})&=&\frac{1}{pp'\sqrt{2pp'}}\sum_{(r',s')\in\Jc}(-1)^{rs'+r's}
  (2-\delta_{r',p})(2-\delta_{s',p'})\nn
 &&\times\Big\{\big[rs\cos\rtt\cos\stt\big]-i\tau\big[r(p'-s')\cos\rtt\sin\stt+(p-r')s\sin\rtt\cos\stt\big]\nn
   &&\quad-\tau^2\big[(p-r')(p'-s')\sin\rtt\sin\stt\big]\Big\}
    \Big(\chit_{r',s'}^+(\tau)+(-1)^{rp'+ps}\chit_{r',s'}^-(\tau)\Big)\nn
 &+&\frac{1}{(pp')^2\sqrt{2pp'}}\sum_{(r',s')\in\Jc_1}(-1)^{rs'+r's}
   \Big\{\big[2rspp'\cos\rtt\cos\stt+\big((rp')^2+(ps)^2\big)\sin\rtt\sin\stt\big]\nn
   &&\quad-i\tau\big[2(r'p'-ps')\big(rp'\cos\rtt\sin\stt-ps\sin\rtt\cos\stt\big)+4\pi pp'\sin\rtt\sin\stt\big]\nn
   &&\quad+\tau^2\big[(r'p'-ps')^2\sin\rtt\sin\stt\big]\Big\}\chit_{r',s'}(\tau) 
\label{Splus}
\eea 
\bea
 \chit_{r,s}^-(-\tfrac{1}{\tau})&=&\frac{1}{pp'\sqrt{2pp'}}\sum_{(r',s')\in\Jc}(-1)^{(p-r)s'+r'(p'-s)}
  (2-\delta_{r',p})(2-\delta_{s',p'})\nn
 &&\times\Big\{\big[rs\cos\rtt\cos\stt\big]-i\tau\big[r(p'-s')\cos\rtt\sin\stt+(p-r')s\sin\rtt\cos\stt\big]\nn
   &&\quad-\tau^2\big[(p-r')(p'-s')\sin\rtt\sin\stt\big]\Big\}
    \Big(\chit_{r',s'}^+(\tau)+(-1)^{rp'+ps+pp'}\chit_{r',s'}^-(\tau)\Big)\nn
 &+&\frac{1}{(pp')^2\sqrt{2pp'}}\sum_{(r',s')\in\Jc_1}(-1)^{(p-r)s'+r'(p'-s)}\nn
 &&\times\Big\{
  \big[2rspp'\cos\rtt\cos\stt+\big((rp')^2+(ps)^2-(pp')^2\big)\sin\rtt\sin\stt\big]\nn
   &&\quad-i\tau\big[2(r'p'-ps')\big(rp'\cos\rtt\sin\stt-ps\sin\rtt\cos\stt\big)+4\pi pp'\sin\rtt\sin\stt\big]\nn
   &&\quad+\tau^2\big[(r'p'-ps')^2\sin\rtt\sin\stt\big]\Big\}\chit_{r',s'}(\tau) 
\eea 
and
\be
 \chit_{r,s}(-\tfrac{1}{\tau})=-\frac{4}{\sqrt{2pp'}}\sum_{(r',s')\in\Jc_1}(-1)^{rs'+r's}
  \sin\rtt\sin\stt\,\chit_{r',s'}(\tau)
\label{Smin}
\ee
These modular transformations are neatly encoded in the $\tau$-dependent $S$-matrix
\be
 \Sc=\sum_{\ell=0}^2(-i\tau)^\ell \Sc^{(\ell)}=\Sc^{(0)}-i\tau \Sc^{(1)}-\tau^2 \Sc^{(2)}
\label{Sc}
\ee
where the three $(2pp'+\tfrac{1}{2}(p-1)(p'-1))$-dimensional and $\tau$-independent
matrices $\Sc^{(\ell)}$ are read off from 
(\ref{Splus})-(\ref{Smin}). We find the inverse of $\Sc$ to be given by
\be
 \Sc^{-1}=\Sc(-1/\tau)=\sum_{\ell=0}^2(-i\tau)^{-\ell}\Sc^{(\ell)}=\Sc^{(0)}+i\tau^{-1}\Sc^{(1)}-\tau^{-2}\Sc^{(2)}
\label{Scinv}
\ee

\subsection{On modular invariants}
\label{AppOnMod}

Let the set of $\Wc$-irreducible characters form the $(2pp'+\tfrac{1}{2}(p-1)(p'-1))$-dimensional vector
\be
 K(\tau)=\left(\!\!\begin{array}{c} \chit^+(\tau) \\ \chit^-(\tau) \\ \chit(\tau) \end{array}\!\!\right)
\label{K}
\ee
with block structure according to the notation of Appendix \ref{AppModular}. 
Modular invariance of the sesquilinear form 
\be
 Z(\tau)=\sum_{i,j=1}^{2pp'+\frac{1}{2}(p-1)(p'-1)} K_i(\tau)M_{ij}K_j(\bar{\tau}),\qquad M_{ij}\in\mathbb{C}
\label{Ztau}
\ee
in these characters thus corresponds to the matrix equations
\be
 \Sc^TM\Sc=M,\qquad \Tc^TM\Tc=M
\ee
where $A^T$ denotes the transpose of $A$.
For later convenience, we introduce the following notation for the block decomposition,
as in (\ref{K}), of the multiplicity matrix $M$ 
\be
 M=\left(\!\!\begin{array}{ccc} 
   M_+^+&M_+^-&M_+^0\\[4pt] 
   M_-^+&M_-^-&M_-^0 \\[4pt] 
   M_0^+&M_0^-&M_0^0\end{array}\!\!\right)
\label{M}
\ee

As usual, and following from the explicit transformation rules worked out in \cite{FGST06},
$T$-invariance implies that the conformal weights of paired characters must differ by integers, that is,
\be
 M_{ij}\neq0\quad \Rightarrow\quad \D_i-\D_j\in\mathbb{Z}
\label{MZ}
\ee
To analyze the implications of $S$-invariance, one can use the decompositions of $\Sc$ and 
its inverse given in (\ref{Sc}) and (\ref{Scinv}). It follows that $S$-invariance of (\ref{Ztau}) is 
equivalent to
\be
 (\Sc^{(1)})^TM=(\Sc^{(2)})^TM=M\Sc^{(1)}=M\Sc^{(2)}=(\Sc^{(0)})^TM-M\Sc^{(0)}=0
\label{SM}
\ee
where it is noted that we do not consider $\tau$-dependent multiplicities.

\subsection{Proof of the proposition in Section \ref{SecSeparation}}
\label{AppProp1}

Our objective here is to prove the proposition in Section~\ref{SecSeparation}.
It is stressed that the present goal is therefore not to classify the modular invariants of the form (\ref{Ztau}),
but to prove that certain characters can appear only in very specific linear combinations. 
Viewing these combinations as conditions on the
space of characters, we thus need to show that they are {\em necessary} conditions for $S$-invariance.
In order to do that, we only need to impose a subset of the conditions in (\ref{SM}), it turns out.
The remaining conditions will restrict the system even further and, combined with (\ref{MZ}), eventually lead to
a classification of the modular invariants.

We begin by considering the nine matrix equations
\be
 [M\Sc^{(1)}]_\mu^\nu=M_\mu^+\Sc_+^{(1)\nu}+M_\mu^-\Sc_-^{(1)\nu}=0,\qquad \mu,\nu=+,-,0
\label{MS1}
\ee
where we have used that $\Sc_0^{(1)\nu}=0$ according to (\ref{Smin}).
In this appendix, column labels are indicated by superscripts.
For every $\mu$, the matrix equation corresponding to a given $\nu$ gives rise to
a set of linear conditions 
\be
 \sum_{r=1}^p\sum_{s=1}^{p'}\Big([M_\mu^+]_{\rho,\sigma}^{r,s}[\Sc_+^{(1)\nu}]_{r,s}^{r',s'}
   +[M_\mu^-]_{\rho,\sigma}^{r,s}[\Sc_-^{(1)\nu}]_{r,s}^{r',s'}\Big)=0
\label{sumM}
\ee
labelled by $(r',s')$, where $(r',s')\in\Jc$ for $\nu=\pm$ while $(r',s')\in\Jc_1$ for $\nu=0$.
Likewise, the domains for $(\rho,\sigma)$ depend on $\mu$.
The conditions (\ref{sumM}) can be written as linear conditions on the combinations
\be
 K_{\mu;\rho,\sigma}^{r,s}=[M_\mu^+]_{\rho,\sigma}^{r,s}-[M_{\mu}^-]_{\rho,\sigma}^{r,p'-s},\qquad
 L_{\mu;\rho,\sigma}^{r,s}=[M_\mu^+]_{\rho,\sigma}^{r,s}-[M_{\mu}^-]_{\rho,\sigma}^{p-r,p'}
\ee
with coefficients given by products of trigonometric functions depending also on $r'$ and $s'$.
{}By manipulating the conditions for $\nu=+$, in particular, we find that they are equivalent to
\be
 \begin{array}{llll} 
 &\displaystyle{K_{\mu;\rho,\sigma}^{r,s}=\frac{p-r}{r}K_{\mu;\rho,\sigma}^{p-r,p'-s}},\qquad 
     &1\leq r\leq p,\quad &1\leq s\leq p'-1
 \\[8pt]
 &\displaystyle{L_{\mu;\rho,\sigma}^{r,s}=\frac{p'-s}{s}L_{\mu;\rho,\sigma}^{p-r,p'-s}},\qquad 
    &1\leq r\leq p-1,\quad &1\leq s\leq p'
 \end{array}
\ee 
Combining these with the $\nu=0$ conditions, we conclude
that (\ref{MS1}) implies that
\be
 [M_\mu^+]_{\rho,\sigma}^{p,s}=[M_\mu^-]_{\rho,\sigma}^{p,p'-s},\quad
 [M_\mu^+]_{\rho,\sigma}^{r,p'}=[M_\mu^-]_{\rho,\sigma}^{p-r,p'},\quad
 [M_\mu^+]_{\rho,\sigma}^{r,s}=[M_\mu^-]_{\rho,\sigma}^{r,p'-s}
 =[M_\mu^-]_{\rho,\sigma}^{p-r,s}=[M_\mu^+]_{\rho,\sigma}^{p-r,p'-s}
\ee
for $1\leq r<p$, $1\leq s<p'$ and general $\mu=\pm,0$.
Likewise, the matrix equations
\be
 [(\Sc^{(1)})^TM]_\nu^\mu=((\Sc^{(1)})^T)_\nu^+M_+^\mu+((\Sc^{(1)})^T)_\nu^-M_-^\mu=0,\qquad \mu,\nu=+,-,0
\label{S1M}
\ee
imply that
\be
 [M^\mu_+]^{\rho,\sigma}_{p,s}=[M^\mu_-]^{\rho,\sigma}_{p,p'-s},\quad
 [M^\mu_+]^{\rho,\sigma}_{r,p'}=[M^\mu_-]^{\rho,\sigma}_{p-r,p'},\quad
 [M^\mu_+]^{\rho,\sigma}_{r,s}=[M^\mu_-]^{\rho,\sigma}_{r,p'-s}
 =[M^\mu_-]^{\rho,\sigma}_{p-r,s}=[M^\mu_+]^{\rho,\sigma}_{p-r,p'-s}
\ee
for $1\leq r<p$, $1\leq s<p'$ and general $\mu=\pm,0$.
It follows that the $S$-invariant (\ref{Ztau}) can be expressed as a sesquilinear form in 
character combinations of the form
\be
 \big\{\chit_{p,p'}^\pm;\    \chit_{p,s}^++\chit_{p,p'-s}^-;\    \chit_{r,p'}^++\chit_{p-r,p'}^-;\   
  \chit_{r,s}^++\chit_{r,p'-s}^-+\chit_{p-r,s}^-+\chit_{p-r,p'-s}^+\big\}\,\cup\,\big\{\chit_{r,s}\big\}
\ee
where $1\leq r<p$ and $1\leq s<p'$, and where we have suppressed the dependence on $\tau$.
This characterization is equivalent to
\be
  \big\{\chit_{p,p'}^\pm;\    \chit_{p,s}^++\chit_{p,p'-s}^-;\    \chit_{r,p'}^++\chit_{p-r,p'}^-;\   
  \tfrac{1}{2}\chit_{r,s}+\chit_{r,s}^++\chit_{r,p'-s}^-+\chit_{p-r,s}^-+\chit_{p-r,p'-s}^+\big\}\,\cup\,
     \big\{\chit_{r,s}\big\}
\ee
thus concluding the proof of the proposition.

%%%%%%%%%%%%%%%%%%%%%%%%%%%%%%%%%%%%%%%%%%%%%%%
%%%%%%%%%%%%%%%%%%%%%%%%%%%%%%%%%%%%%%%%%%%%%%%

\end{document}